\documentclass[amsmath,amssymb,14pt]{article}

\usepackage{graphicx, epsfig}

\usepackage{amssymb,amsmath,color}
\usepackage{amsthm}
\usepackage[a4paper, total={6.5in, 10in}]{geometry}

\numberwithin{equation}{section}

\usepackage{subcaption}
\usepackage{caption}
\usepackage[export]{adjustbox}

\usepackage[nottoc]{tocbibind}

\usepackage{dcolumn}
\usepackage{bm}

\newtheorem{mytheo}{Theorem}[section]
\newtheorem{mylemma}{Lemma}[section]
\newtheorem{mycorollary}[mytheo]{Corollary}
\newtheorem{myproposition}[mytheo]{Proposition}

\newtheorem{myremark}[mytheo]{Remark}

\newtheorem{mytheo*}{Theorem}
\newtheorem{mylemma*}{Lemma}
\newtheorem{myproposition*}{Proposition}
\newtheorem{mydef*}{Definition}[section]
\newtheorem{myremark*}{Remark}
\newtheorem{myproof*}{Proof}
\newtheorem*{mynotation*}{Notation}
\newtheorem{mycorollary*}{Corollary}
\newtheorem*{theorem*}{Theorem}

\usepackage{xcolor}

\DeclareMathOperator{\Tr}{Tr}
\DeclareMathOperator{\Imaginary}{Im}
\DeclareMathOperator{\Real}{Re}
\DeclareMathOperator{\Ran}{Ran}
\DeclareMathOperator{\Ker}{Ker}
\DeclareMathOperator{\ci}{ci}
\DeclareMathOperator{\si}{si}

\usepackage{mathtools}

\usepackage{authblk}

\begin{document}
\title{\textsc{Rank One Perturbations Supported by Hybrid Geometries and Their Deformations}}

\author[1]{Fatih Erman}
\author[2]{Sema Seymen}
\author[3]{O. Teoman Turgut}
\affil[1]{Department of Mathematics, \.{I}zmir Institute of Technology, Urla, 35430, \.{I}zmir, Turkey}
\affil[2, 3]{Department of Physics, Bo\u{g}azi\c{c}i University, Bebek, 34342, \.{I}stanbul, Turkey}
\affil[1]{fatih.erman@gmail.com}
\affil[2]{seymen@itu.edu.tr}
\affil[3]{turgutte@boun.edu.tr}

\maketitle

\begin{abstract}
We study the hybrid type of rank one perturbations in $\mathbb{R}^2$ and $\mathbb{R}^3$, where the perturbation supported by a circle/sphere is  considered together with the delta potential supported by a point outside of the circle/sphere. The construction of the self-adjoint Hamiltonian operator associated with the formal expressions for the rank one perturbation supported by a circle and by a point is explicitly given. The bound state energies and scattering properties for each problem are also studied. Finally, we consider the rank one perturbation supported by a deformed circle/sphere and show that the first order change in the bound state energies under small deformations of the circle/sphere has a simple geometric interpretation.
\end{abstract}

Keywords: Rank one perturbation supported by curves and surfaces, resolvent, Krein's resolvent formula, bound states, scattering. 

%%%%%%%%%%%%%%%%%%%%%%%%%%%%%%%%%%%%%%%%%%%%%%%%%%%%%%%%%
\section{Introduction}

Dirac delta potentials, also known as point interactions in general, are among the exactly solvable classes of potentials studied from both physical and mathematical points of view. A detailed review  of the subject together with their mathematically rigorous constructions as well as  spectral properties are given in the monographs  \cite{Albeverio2012solvable, AlbeverioKurasov}.  

Extension of point delta potentials to the ones whose support is  a sphere - known as the delta shell potentials - is formally presented in some quantum mechanics textbooks  \cite{Demkov, Gottfried, Griffiths}, but its first precise mathematical treatment is given in \cite{AntoineGesztesyShabani} to the best of our knowledge. In that work,  an additional  delta potential supported by a point at the center of the sphere is also discussed in the zero angular momentum sector $l=0$. Such circular/spherical singular interactions, also considered to be models for circular/spherical quantum billiards are studied analytically and numerically recently \cite{maioli2018exact, azado2021quantum}. Their higher dimensional generalization has been worked out in the literature from the point of view of differential equations \cite{Demiralp2003properties} using the partial wave analysis. In addition to these specific geometrical objects as the supports of delta function, more complicated contact type interactions are  studied by considering some regular curves or surfaces. The Schr\"{o}dinger operators or Hamiltonians for such class of singular interactions are formally given by
\begin{eqnarray}
H_{\Gamma} & = & H_0 - \lambda \delta (\cdot- \Gamma) \;, \\ H_{\Sigma} & = & H_0  - \lambda \delta (\cdot- \Sigma) \;,
\end{eqnarray}
respectively. Here $H_0$ is the free Hamiltonian. There are various ways to define the above Hamiltonians in a mathematically proper way. If the curve is planar, similarly if  the surface is embedded in $\mathbb{R}^3$, (with some regularity conditions on the curves/surfaces), one  way is to interpret these interactions expressed by the following quadratic forms $\int_{\mathbb{R}^2} |\nabla \psi|^2 d^2 x - \lambda \int_{\Gamma} |\psi|^2 d s \;,$ and $\int_{\mathbb{R}^3} |\nabla \psi|^2 d^3 x - \lambda \int_{\Sigma} |\psi|^2 d A$, respectively and then prove that there exist associated self-adjoint Hamiltonians \cite{brasche1994schrodinger, exner2002bound, ExnerYoshitomi2002, ExnerYoshitomi2003, Kondej2016}. Other  ways are to impose proper  boundary conditions (continuity and jump discontinuity condition at $\Gamma$, see Remark 4.1 in \cite{brasche1994schrodinger} and \cite{exner2003spectra}) or employ scaled potentials \cite{exner2001geometrically}, or direct construction of the resolvent \cite{brasche1994schrodinger, ExnerFraas09, posilicano2001krein} (see also \cite{exner2002curvature}, where the support of the delta potential is a curve in three dimensions). The main physical motivation for studying such systems is to give a realistic model for trapped electrons due to interfaces between two different semiconductor materials, which are known as leaky graphs, curves or surfaces in the literature \cite{Exner}. Small geometric deformations of the support of the delta potentials recently attract some attention \cite{ExnerFraas09}, where the area preserving small deformations can give rise to the isolated eigenvalues. Furthermore, the scattering theory for delta-potentials supported by locally deformed planes is constructed in \cite{CacciapuotiFermiPosilicano}. The generalizations to delta functions supported on curves and surfaces embedded in manifolds are presented in the works \cite{burak1, burak2}.

The aim of this paper is to address the issue of the bound state and scattering spectrum of the free Hamiltonian operators perturbed by the following rank one perturbations:
\begin{itemize}
    \item[(i)] Rank one perturbation supported by a circle and by a point outside of the circle in $\mathbb{R}^2$, 
    \item[(ii)] Rank one perturbation supported by a sphere and by a point outside of the sphere in $\mathbb{R}^3$,
    \item[(iii)] Rank one perturbation supported by a small deformation in the normal direction of a circle in $\mathbb{R}^2$,
    \item[(iv)] Rank one perturbation supported by a small deformation in the normal direction of a sphere in $\mathbb{R}^3$.
\end{itemize}
Actually, such perturbations are singular rank one perturbations in the sense described in \cite{AlbeverioKurasov} but we will simply call them rank one perturbations throughout this paper. For instance, the Hamiltonian for the first problem is formally given by 
\begin{eqnarray}
H_0 -\lambda_1 |\mathbf{a} \rangle \langle \mathbf{a}| - \lambda_2 |\Gamma \rangle \langle \Gamma | \;, \label{formalHamiltoniancirclepoint}
\end{eqnarray}
where $\mathbf{a} \in \mathbb{R}^2$, $\Gamma$ is the circle centered at the origin with radius $R$ in $\mathbb{R}^2$, and $\lambda_1, \lambda_2>0$.
\textit{In this paper, we shall call  discrete eigenvalues of the Hamiltonian below its essential spectrum the bound state energies, which is a common terminology in physics.} %

It is well known that the resolvent of such singular interactions (delta potential supported by a point, a curve, or a surface) can be expressed by some explicit formulae involving the resolvent of the free Hamiltonian. These expressions are commonly known as Krein's formula in the literature \cite{Albeverio2012solvable, AlbeverioKurasov, Exner}. To find the resolvent of the above hybrid type of potentials, we essentially follow an approach  given in \cite{rajeevdimock}. For this, we first regularize the ill-defined interaction terms by finite rank projections acting on Hilbert space and then find the resolvent associated with these regularized Hamiltonians. Then, considering the strong limit of these resolvents, as we remove the regularization parameter,  allows us to define a self-adjoint operator corresponding to  their limits thanks to the Trotter-Kato theorem. If the support of the interaction is co-dimension two or three, then it is well-known that we need to renormalize the problem (see e.g., \cite{Huang, Jackiw} for the point interactions in two and three dimensions). In this case, we need to choose the coupling constants or strengths as functions of the regularization parameter such that the limit converges. It is important to emphasize that the second term in the above Hamiltonian (\ref{formalHamiltoniancirclepoint}) defines a (singular) rank one perturbation supported on the circle. It turns out,  the Hamiltonian in equation (\ref{formalHamiltoniancirclepoint}) corresponds to the  zero angular momentum sector of delta potentials supported on a circle (Similar models  together with some regular potentials have been studied in \cite{Behrndt2013} using quasi boundary triples. For an alternative treatment of scattering  applied to  co-existing point and line defects, see also the recent work \cite{mostafazadeh}).

For the sake of brevity, we only present the construction of the self-adjoint operator associated with the first system (i) and skip  the technical details of the construction of the self-adjoint Hamiltonian associated with the other systems (ii)-(iv) since the idea of
construction is essentially the same. One of the main results of the paper is to give explicitly the bound state energies and differential cross sections for  systems (i)-(ii). The physical motivation of considering systems in (i) and (ii) is based on the extension of previously leaky curves/surfaces described in \cite{Exner} to the case where impurities in the semiconductor can be modeled by point delta interactions in the low energy approximation (we restrict the problem to the $l=0$ sector) for simplicity. Moreover, we  show that the change in the bound state energies under small deformations in the normal directions of the circle/sphere can be similarly  studied to the first order in the deformation.
An interesting observation here is that  the first order  perturbative calculation of the bound state energy  gives the same result as  the rank one perturbation supported by a circle/sphere with a radius increased by the  average of the deformation. The method developed in this paper is in fact rather general and can be applied also to rank one perturbation supported by curves and surfaces in principle.

 The paper is organized as follows. In Section \ref{Delta Potential Supported by a Circle and a Point}, we explicitly show that there exists a self-adjoint operator associated with the initial formal Hamiltonian where the interaction contains a rank one perturbation supported by a circle centered at the origin and a point outside of this circle (point being inside does not present any difficulties, it can equally be considered). Then, we briefly discuss the bound state analysis as well as  scattering solutions. Section \ref{Delta Potential Supported by a Sphere and a Point} deals with the bound state spectrum and scattering  states for the rank one perturbation supported by a sphere centered at the origin and a point outside of the sphere. Moreover, we study how the small deformations of the circle and sphere in the normal directions change the bound state spectrum and scattering properties in Sections \ref{Small Deformations of a Circle} and \ref{Small Deformations of a Sphere}. Finally, Appendix is devoted to the Trotter-Kato theorem, which is needed to prove the self-adjointness of the Hamiltonian.

\begin{mynotation*}
In our formulae we often use the Dirac notation for the inner products, however  this notation is particularly designed for self-adjoint operators (often the distinction between symmetric and self-adjoint is ignored in physics literature), here we need to keep in mind that the operators appearing in our expressions always act on the right,  unless specified otherwise. We shall also use the notation $*$ and $\dagger$ for the complex conjugation of a complex number and adjoint of an operator, respectively. 
Dirac delta function supported by a point $\mathbf{a}$ is defined on the test functions $\psi$ by $\langle \delta_{\mathbf{a}}| \psi \rangle =\langle \mathbf{a}| \psi \rangle := \psi(\mathbf{a})$. Similarly, the Dirac delta function $\delta_{\Gamma}$ supported by the curve $\Gamma$ and the Dirac delta function $\delta_{\Sigma}$ supported by the surface $\Sigma$ are defined by their action on $\psi$ \cite{Appel2007mathematics}
\begin{eqnarray}
\langle \delta_{\Gamma}| \psi \rangle = \langle \Gamma | \psi \rangle & := & \frac{1}{L(\Gamma)} \int_{\Gamma} \psi \; d s \;, \\   \langle \delta_{\Sigma}| \psi \rangle = \langle \Sigma | \psi \rangle & := & \frac{1}{A(\Sigma)} \int_{\Sigma} \psi \; d A \;,  
\end{eqnarray}
where $ds$ is the integration element over the curve $\Gamma$ and $d A$ is the integration element over the surface. For the circle $\Gamma=S^1$, $ds=R d \theta$ and $L(\Gamma)=2 \pi R$. For the sphere $\Sigma=S^2$, $d A =R^2 \sin \theta d \theta d \phi$ and $A(\Sigma)=4 \pi R^2$. The volume elements in $\mathbb{R}^n$ are denoted by $d^n r$.The formal expressions $|\mathbf{a} \rangle \langle \mathbf{a}|$, $|\Gamma \rangle \langle \Gamma|$, and $|\Sigma \rangle \langle \Sigma|$ are also written as $\langle \delta_{\mathbf{a}}, \cdot \rangle \delta_{\mathbf{a}}$, $\langle \delta_{\Gamma}, \cdot \rangle \delta_{\Gamma}$, and $\langle \delta_{\Sigma}, \cdot \rangle \delta_{\Sigma}$, respectively in the literature.

\end{mynotation*}

\section{Rank One Perturbation Supported by a Circle and a Point in $\mathbb{R}^2$}
\label{Delta Potential Supported by a Circle and a Point}

We first consider the rank one perturbation supported by a circle and a point, given formally in Dirac's notation by 
\begin{eqnarray}
   H_{a \Gamma}  =  H_0 -\lambda_1 |\mathbf{a} \rangle \langle \mathbf{a}| - \lambda_2 |\Gamma \rangle \langle \Gamma | \;,
\end{eqnarray}
where $\Gamma$ is the circle centered at the origin with radius $R$ and $D(H_0)=H^2(\mathbb{R}^2)$. We shall use units such that $\hbar=2m=1$ for simplicity. In order to make sense of the above expression, we first regularize the Hamiltonian $H_{a \Gamma}$ by heat kernel $K_{\epsilon/2}$ in the following way:
\begin{eqnarray} \label{regularizedH1}
   H_{a \Gamma, \,\epsilon}= H_0 - \lambda_1(\epsilon) |\mathbf{a}^{\epsilon}\rangle \langle \mathbf{a}^{\epsilon}| - \lambda_2 |\Gamma^{\epsilon} \rangle \langle \Gamma^{\epsilon} | \;,
\end{eqnarray}
where 
\begin{eqnarray}
    \langle \mathbf{a}^{\epsilon}|\psi \rangle & = &  \int_{\mathbb{R}^2} K_{\epsilon/2}(\mathbf{r}, \mathbf{a}) \psi(\mathbf{r}) \; d^2 r \;, \label{definitionaepsgammaeps} \\ 
     \langle \Gamma^{\epsilon} |\psi \rangle & = & \frac{1}{L(S^1)} \int_{S^1} \left(\int_{\mathbb{R}^2}   K_{\epsilon/2}(\mathbf{r}, \boldsymbol{\gamma}(s)) \psi(\mathbf{r}) \; d^2 r \right) \; d s \;. \label{definitionaepsgammaeps2}
\end{eqnarray}
Here $\epsilon>0$ is the regularization parameter or cut-off and $\boldsymbol{\gamma}(s)=(R\cos (s/R), R \sin (s/R))$ is the parametrization of the circle $S^1$. The explicit form of the heat kernel in $\mathbb{R}^n$ is given by \cite{Evans}
\begin{eqnarray} \label{heatkernel}
    K_{\epsilon/2}(\mathbf{r}, \mathbf{r}')= \frac{1}{(2 \pi \epsilon)^{n/2}} \; e^{-|\mathbf{r}- \mathbf{r}'|^2/2\epsilon} \;.
\end{eqnarray}
The strength or coupling constant of the point Dirac delta interaction, denoted by $\lambda_1$, is considered to be a function of $\epsilon>0$, whose explicit form will be determined later. 
In this way, $H_{a \Gamma, \,\epsilon}$ becomes a finite rank perturbation of the free Hamiltonian so that it is self-adjoint on the domain of $H_0$ thanks to the Kato-Rellich theorem \cite{Reedsimonv2}. This choice of the regularization is based on the fact that the heat kernel converges to the Dirac delta function as $\epsilon \rightarrow 0^+$ in the distributional sense. This is an especially natural choice if we consider delta potentials in manifolds \cite{pointinteractionsonmanifolds1, pointinteractionsonmanifolds2} and here it is not only useful for the regularization but also allows us to approximate a singular interaction supported by a circle with a more regular one. It is important to emphasize that the expressions (\ref{definitionaepsgammaeps}) and (\ref{definitionaepsgammaeps2}) are well defined for functions $\psi$ in $H^2(\mathbb{R}^2)$ using Cauchy-Schwarz inequality and reproducing property of the heat kernel $\int_{\mathbb{R}^2} K_{\epsilon/2}(\mathbf{r_1},\mathbf{r}) K_{\epsilon/2}(\mathbf{r},\mathbf{r_2}) d^2 r = K_{\epsilon}(\mathbf{r}_1,\mathbf{r}_2)$. For simplicity, Let us define $|f_1(\epsilon) \rangle =|\mathbf{a}^{\epsilon} \rangle $ and  $|f_2(\epsilon) \rangle= |\Gamma^{\epsilon} \rangle $. Then, we have the following first result:
\begin{myproposition} \label{proposition1}Let $\lambda_1(\epsilon)$ be a continuous function of $\epsilon$, which converges to zero as $\epsilon \to 0^+$ and $\lambda_2> 0$ be an arbitrary positive real number. The resolvent of the regularized Hamiltonian 
\begin{eqnarray}
H_{a \Gamma, \, \epsilon} =H_0- \sum_{i=1}^{2} \lambda_i(\epsilon) |f_i(\epsilon) \rangle \langle f_i(\epsilon)|
\end{eqnarray}
is given by
\begin{eqnarray} \label{regularizedresolventcirclepoint}
    R(\epsilon, E)= R_0(E) + R_0(E) \sum_{i,j=1}^{2} | f_i (\epsilon)\rangle \left(\Phi^{-1}(\epsilon, E) \right)_{ij} \langle f_j(\epsilon)|  R_0(E) \;,
\end{eqnarray}
where 
\begin{eqnarray}
    \Phi_{ij}(\epsilon, E)= 
      \frac{\delta_{ij}}{\lambda_{i}(\epsilon)} - \langle f_i(\epsilon)| R_0(E) |f_j(\epsilon) \rangle \;,
\end{eqnarray}
and its resolvent set is given by $\rho(H_{a \Gamma, \, \epsilon})=\{E \in \rho(H_0): \det(\Phi(\epsilon, E)) \neq 0 \; \text{for} \; \text{all} \;  \epsilon>0 \}$.
\end{myproposition}

\begin{myremark}
As emphasized in the introduction, here  the Dirac's notation $\langle \tilde{f}_i(\epsilon)| R_0(E) | \tilde{f}_i(\epsilon) \rangle $ should be interpreted as  $\langle \tilde{f}_i(\epsilon)| R_0(E) \tilde{f}_i(\epsilon) \rangle$. 
\end{myremark}

\begin{proof} To find the resolvent of the regularized Hamiltonian (\ref{regularizedH1}), we need to solve the following inhomogenous Schr\"{o}dinger equation
\begin{eqnarray}
    (H_{a \Gamma, \, \epsilon}-E)|\psi \rangle = |\rho \rangle \;, \label{inhomogenoussch1}
\end{eqnarray}
for a given function $\langle \mathbf{r}|\rho \rangle= \rho(\mathbf{r}) \in L^2(\mathbb{R}^2)$. The existence of the solution is guaranteed by the basic self-adjointness criteria $\Ran(H_{a \Gamma, \, \epsilon}-E)=L^2(\mathbb{R}^2)$ for at least one $E$ in the upper half-plane and one in the lower half-plane \cite{Reedsimonv2}. It is useful to express the interaction as the sum of the rescaled projection operators:
\begin{eqnarray}
    H_{a \Gamma, \, \epsilon} =H_0 - \sum_{j=1}^{2} |\tilde{f}_j(\epsilon) \rangle \langle \tilde{f}_j(\epsilon)| \;,
\end{eqnarray}
where $|\tilde{f}_i(\epsilon) \rangle = \sqrt{\lambda_i(\epsilon)}|f_i(\epsilon) \rangle$ in $L^2(\mathbb{R}^2)$.
Then, applying the free resolvent $R_0(E)=(H_0-E)^{-1}$ defined on the resolvent set $\rho(H_0)=\mathbb{C} \setminus [0,\infty)$ to the equation (\ref{inhomogenoussch1}), we find 
\begin{eqnarray}
    |\psi \rangle = R_0(E) |\rho \rangle + R_0(E)  \sum_{j=1}^{2} |\tilde{f}_j(\epsilon) \rangle \langle \tilde{f}_j(\epsilon)| \psi \rangle \;. \label{psimodel1unknown}
\end{eqnarray}
The right hand side of this expression involves unknown complex numbers $\langle \tilde{f}_j(\epsilon)| \psi \rangle $. In order to find them, we project this equation onto the $\langle \tilde{f}_i(\epsilon)|$ and isolate the $j=i$th term in the summation to get the following matrix equation
\begin{eqnarray}
    \sum_{j=1}^{2} \tilde{\Phi}_{ij}(\epsilon, E) \langle \tilde{f}_j(\epsilon)|\psi \rangle = \langle \tilde{f}_i(\epsilon) |R_0(E)|\rho \rangle \;, \label{phimatrixequation}
\end{eqnarray}
where
\begin{eqnarray}
\tilde{\Phi}_{ij} (\epsilon, E) =   \begin{cases} 
      1- \langle \tilde{f}_i(\epsilon)| R_0(E) | \tilde{f}_i(\epsilon) \rangle  & i=j \\
      - \langle \tilde{f}_i(\epsilon)| R_0(E) | \tilde{f}_j(\epsilon) \rangle & i \neq j \;.
      \end{cases}
\end{eqnarray}
Assume that the matrix $\tilde{\Phi}$ is invertible for some subset of the free resolvent set to be determined below. Then, the solution of (\ref{phimatrixequation}) exists and is unique. Substituting this solution into (\ref{psimodel1unknown}), we get
\begin{align}
    |\psi \rangle = R_0(E) |\rho \rangle + R_0(E) \sum_{i,j=1}^{2} | \tilde{f}_i \rangle \left(\tilde{\Phi}^{-1}(\epsilon, E) \right)_{ij} \langle \tilde{f}_j|  R_0(E)|\rho \rangle   \;.
\end{align}
Then, the resolvent of the regularized Hamiltonian can be directly read from the above result
\begin{eqnarray}
    R(\epsilon, E)= R_0(E) + R_0(E) \sum_{i,j=1}^{2} | \tilde{f}_i (\epsilon)\rangle \left(\tilde{\Phi}^{-1}(\epsilon, E) \right)_{ij} \langle \tilde{f}_j(\epsilon)|  R_0(E) \;.
\end{eqnarray}
It is convenient to express the above sum in the following way 
\begin{eqnarray}
   \sum_{i,j=1}^{2} | \tilde{f}_i (\epsilon) \rangle \left(\tilde{\Phi}^{-1}(\epsilon, E) \right)_{ij} \langle \tilde{f}_j(\epsilon)|  & = & \Tr\left(\tilde{F}(\epsilon) \tilde{\Phi}^{-1}(\epsilon, E)\right) \;, \label{regularizedresolvent1ststep}
\end{eqnarray}
where we have defined the matrix $\tilde{F}_{ij} :=  | \tilde{f}_i \rangle \langle \tilde{f}_j|$. If we define the diagonal matrix $D_{ij}(\epsilon):= \sqrt{\lambda_{i}(\epsilon)} \delta_{ij}$ we can decompose $\tilde{F}=D F D$, where $F_{ij}=|f_i\rangle \langle f_j|$. This helps us to write the summation term (\ref{regularizedresolvent1ststep}) as $\Tr\left(\tilde{F} \tilde{\Phi}^{-1}\right) = \Tr(D F D \tilde{\Phi}^{-1}) = \Tr(F D \tilde{\Phi}^{-1} D) = \Tr\left(F \Phi^{-1}\right)$, where $\Phi$ is related to $\tilde{\Phi}$ by a similarity transformation $\Phi=D^{-1} \tilde{\Phi} D^{-1}$ and given by
\begin{eqnarray} \label{regularizedPhi1}
    \Phi_{ij}(\epsilon, E)= 
    \begin{cases} 
      \frac{1}{\lambda_{i}(\epsilon)} - \langle f_i(\epsilon)| R_0(E) |f_i(\epsilon) \rangle  & i=j \\
      - \langle f_i(\epsilon)| R_0(E) |f_j(\epsilon) \rangle & i \neq j \;.
      \end{cases} 
\end{eqnarray}
Hence, we explicitly find the resolvent formula for the regularized Hamiltonian (\ref{regularizedresolventcirclepoint}).
We now show that $E \in \rho(H_0)$ lies in the resolvent set for $(H_{a \Gamma, \, \epsilon}-E)$ if and only if the matrix $\Phi(\epsilon, E)$ is invertible. To prove this, we first assume that   $\Phi(\epsilon, E)$ is invertible for some values of $E \in \rho(H_0)$. From the triangle inequality, we have
\begin{eqnarray}
    ||R(\epsilon, E)|\psi \rangle|| \leq ||R_0(E)|\psi \rangle || + 4 \max_{1\leq i,j \leq 2} |\left(\Phi^{-1}(\epsilon, E) \right)_{ij}| \; |\langle f_j(\epsilon)|  R_0(E)|\psi \rangle| \; ||R_0(E) | f_i (\epsilon)\rangle|| \;. \label{Rpsibound}
\end{eqnarray}
We need to show that the right hand side of this inequality is a bounded function of $E$ where $E$ must lie in $\rho(H_0)$ and satisfy $\det \Phi(\epsilon, E) \neq 0$. Moreover, this bound must also be a regular function of $\epsilon$ since we will consider the limiting case as $\epsilon \to 0^+$ by appropriately choosing $\lambda_1(\epsilon)$, as we will show later on.

A direct application of Cauchy-Schwarz inequality to the inner product in the right hand side of the above inequality does not yield a regular estimate in $\epsilon$ since the norm of the function $f_i(\epsilon)$ is not regular. For this reason, we may think that the adjoint of the  bounded free resolvent operator acts on the first entry in the inner product and then apply the Cauchy-Schwarz inequality 
\begin{eqnarray}
    |\langle f_i(\epsilon)|  R_0(E)|\psi \rangle| \leq ||R_0(E^*)|f_i(\epsilon)\rangle|| \,||\psi||< \infty \;,
    \end{eqnarray}
where we have used the fact that $R_0^\dag (E)=R_0(E^*)$. Since $E$ is inside the resolvent set of $H_0$, the expression $||R_0(E) | f_i (\epsilon)\rangle||$ and the inner product on the right hand side of the inequality (\ref{Rpsibound}) is finite as long as $|f_i(\epsilon) \rangle$ lies in $L^2(\mathbb{R}^2)$. However, we must also show that their bounds must be regular in $\epsilon$ as $\epsilon \to 0^+$.  It is easy to see that
\begin{eqnarray}
||R_0(E^*)|\mathbf{a}^{\epsilon} \rangle||^2 = \int_{\mathbb{R}^2} \frac{|\langle \mathbf{p}|\mathbf{a}^{\epsilon} \rangle|^2}{(p^2-E)(p^2 - E^*)} \; \frac{d^2 p}{(2\pi)^2} \;. \label{normsquareofR0a}
\end{eqnarray}
Using (\ref{definitionaepsgammaeps}) and the explicit form of the heat kernel (\ref{heatkernel}), we find
\begin{eqnarray}
\langle \mathbf{p}|\mathbf{a}^{\epsilon} \rangle = \frac{e^{-i \mathbf{p} \cdot \mathbf{a}}}{2\pi \epsilon} \int_{\mathbb{R}^2} e^{-i \mathbf{p} \cdot (\mathbf{r}-\mathbf{a})} e^{-\frac{|\mathbf{r}-\mathbf{a}|^2}{2\epsilon}}d^2 r \;.
\end{eqnarray}
By writing the integral in polar coordinates and using the integral representation of the Bessel function of the first kind $J_0(x)$
\begin{eqnarray}
J_0(x)=\frac{1}{2\pi} \int_{0}^{2\pi} e^{-i x \cos \theta} \; d \theta \label{intrepofbessel1stkind}
\end{eqnarray}
and the formula \cite{gradshteyn2014table}
\begin{eqnarray}
\int_{0}^{\infty} x^{\nu +1} e^{-\alpha x^2} J_{\nu}(\beta x) \; \; d x = \frac{\beta^{\nu}}{(2\alpha)^{\nu+1}} \; e^{-\frac{\beta^2}{4 \alpha}} \;, 
\end{eqnarray}
we get
\begin{eqnarray}
\langle \mathbf{p}|\mathbf{a}^{\epsilon} \rangle = e^{- i \mathbf{p} \cdot \mathbf{a}} \; e^{-\epsilon p^2 /2} \;. \label{paepsilon}
\end{eqnarray}
Substituting this result into (\ref{normsquareofR0a}) yields the following bound
\begin{eqnarray}
||R_0(E^*)|\mathbf{a}^{\epsilon} \rangle||^2  \leq \frac{1}{2\pi} \int_{0}^{\infty} \frac{e^{-\epsilon p^2} p}{|p^4 -2 p^2 \Real(E) + (\Real(E)^2 + \Imaginary(E)^2)|} \; d p \;.
\end{eqnarray}
Except for the positive real $E$ axis, the above integral converges and one can estimate its upper bound if $\Real(E)<0$ by 
\begin{eqnarray}
 ||R_0(E^*)|\mathbf{a}^{\epsilon} \rangle||^2  \leq \frac{1}{2\pi} \int_{0}^{\infty} \frac{e^{-\epsilon p^2} p}{p^4 + A} \; d p \;,
\end{eqnarray}
where $A=\Real(E)^2 + \Imaginary(E)^2$. Thanks to the result (3.354) given in \cite{gradshteyn2014table}, we can evaluate the above integral so that
\begin{eqnarray}
 ||R_0(E^*)|\mathbf{a}^{\epsilon} \rangle||^2  \leq \frac{1}{4\pi \sqrt{A}} \left( \ci(\epsilon \sqrt{A}) \sin(\epsilon \sqrt{A}) - \si(\epsilon \sqrt{A}) \cos(\epsilon \sqrt{A}) \right)  \;, \label{R0abound}
\end{eqnarray}
where $\si(x)=-\int_{x}^{\infty} \frac{\sin t}{t} dt$ is the sine integral function, and $\ci(x)=-\int_{x}^{\infty} \frac{\cos t}{t} dt$ is the cosine integral function. It is easy to see that this bound is a regular function of $\epsilon$ for all $A\neq 0$.

If $\Real(E)>0$, 
\begin{eqnarray}
||R_0(E^*)|\mathbf{a}^{\epsilon} \rangle||^2 & = &  \frac{1}{4\pi} \int_{0}^{\infty} \frac{e^{-\epsilon u}}{(u-\Real(E))^2+\Imaginary(E)^2} \; d u \nonumber \\
& = & \frac{1}{4\pi} \int_{-\Real(E)}^{0} \frac{e^{-\epsilon (v+\Real(E))}}{v^2+\Imaginary(E)^2} \; d v + \frac{1}{4\pi} \int_{0}^{\infty} \frac{e^{-\epsilon (v+\Real(E))}}{v^2+\Imaginary(E)^2} \; d v \nonumber \\ & \leq & \frac{1}{4\pi} \int_{-\Real(E)}^{0} \frac{1}{v^2+\Imaginary(E)^2} \; d v + \frac{1}{4\pi} \int_{0}^{\infty} \frac{e^{-\epsilon v}}{v^2+\Imaginary(E)^2} \; d v \; , \label{R0abound2}
\end{eqnarray}
which are finite and regular in $\epsilon$ by the same reason given above. We can similarly show that the following norm 
\begin{eqnarray}
||R_0(E^*)|\Gamma^{\epsilon} \rangle||^2 = \int_{\mathbb{R}^2} \frac{|\langle \mathbf{p}|\Gamma^{\epsilon} \rangle|^2}{(p^2-E)(p^2 - E^*)} \; \frac{d^2 p}{(2\pi)^2} \;, \label{normsquareofR0g}
\end{eqnarray}
is a bounded function of $E$ on $\rho(H_0)$ and regular in $\epsilon$. For this, we need to find 
\begin{eqnarray}
\langle \mathbf{p} | \Gamma^{\epsilon} \rangle = \frac{1}{L} \int_{\mathbb{R}^2} e^{i \mathbf{p} \cdot \mathbf{r}} \left( \int_{S^1} K_{\epsilon/2}(\mathbf{r}, \boldsymbol{\gamma}(s)) \; d s \right) \; d^2 r \;.
\end{eqnarray}
Using the explicit expression of the heat kernel (\ref{heatkernel}) and the integral representation of the modified Bessel function of the first kind \cite{Lebedev1965special}
\begin{eqnarray}
    I_0(x)= \frac{1}{2\pi} \int_{0}^{2\pi} e^{x \cos \theta} d\theta \label{intrepI0} \;,
\end{eqnarray}
we get
\begin{eqnarray}
\int_{S^1} K_{\epsilon/2}(\mathbf{r}, \boldsymbol{\gamma}(s)) \; d s = \frac{R}{\epsilon} \; e^{-\frac{(r^2 + R^2)}{2\epsilon}} \; I_0 \left(\frac{R}{\epsilon}r\right) \;.
\end{eqnarray}
Then, from the result (6.633) in \cite{gradshteyn2014table}
\begin{eqnarray}
\int_{0}^{\infty} x \, e^{-\alpha x^2} I_{\nu}(\beta x) J_{\nu}(\gamma x) \; d x = \frac{1}{2\alpha} e^{\frac{(\beta^2-\gamma^2)}{4 \alpha}} J_{\nu}\left(\frac{\beta \gamma}{2 \alpha}\right) \;,
\end{eqnarray}
we obtain
\begin{eqnarray}
\langle \mathbf{p} | \Gamma^{\epsilon} \rangle = e^{-\frac{\epsilon p^2}{2}} \, J_0(p R) \;. \label{pgammaepsilon}
\end{eqnarray}
Combining all these results yield 
\begin{eqnarray}
||R_0(E^*)|\Gamma^{\epsilon} \rangle||^2 = \frac{1}{2\pi} \int_{0}^{\infty} \frac{p \, e^{-\epsilon p^2} J_{0}^{2}(p R)}{(p^2-E)(p^2-E^*)} \; d p \;.
\end{eqnarray}
Since $J_{0}^{2}(p R) \leq 1$, we obtain the same form of the estimate (\ref{R0abound}) and (\ref{R0abound2}) as above.
All these imply that $R(\epsilon,E)$ is bounded  if $E \in \rho(H_0)$ and satisfies $\det(\Phi(\epsilon,E)) \neq 0$, hence $E \in \rho(H_{a \Gamma, \, \epsilon})$.

Conversely, if $E \in \rho(H_{a \Gamma, \, \epsilon})$, then $\det(\Phi(\epsilon,E)) \neq 0$. For this, suppose that $E \in \mathbb{C}\setminus [0, \infty)$ satisfies $\det(\Phi(\epsilon,E))=0$. We need to show that $E \notin \rho(H_{a \Gamma, \, \epsilon})$ or $E$ lies in the spectrum of $H_{a \Gamma, \, \epsilon}$, or in particular $E$ is an eigenvalue of $H_{a \Gamma, \, \epsilon}$:
\begin{eqnarray}
    H_{a \Gamma, \, \epsilon} |\psi \rangle = E |\psi \rangle \;, \label{regularizedeigenvalueequation}
\end{eqnarray}
for some non-zero $|\psi \rangle \in L^2(\mathbb{R}^2)$. The above eigenvalue problem for the regularized Hamiltonian is equivalent to the problem of finding non-trivial solution $\langle \tilde{f}_j(\epsilon)|\psi \rangle$ of the equation (\ref{phimatrixequation}) with $|\rho\rangle=|0\rangle$:
\begin{eqnarray}
\sum_{j=1}^{2} \tilde{\Phi}_{ij}(\epsilon,E) \langle \tilde{f}_j(\epsilon)|\psi \rangle =0 \;. \label{eigenvalueequation2ndform}
\end{eqnarray}
Since the equation (\ref{eigenvalueequation2ndform}) is derived from the eigenvalue equation (\ref{regularizedeigenvalueequation}) of the regularized Hamiltonian, the set of $E$ in (\ref{regularizedeigenvalueequation}) must also satisfy the equation (\ref{eigenvalueequation2ndform}). To prove the converse, we first need to show that $\langle f_i(\epsilon)| R_0(E) |f_j(\epsilon) \rangle \neq 0$ for all $i,j$. Otherwise, $\tilde{\Phi}$ would be an identity matrix, which is clearly invertible. Then, the equation (\ref{eigenvalueequation2ndform}) implies that $\langle \tilde{f}_j(\epsilon)|\psi \rangle=0$ for all $j$. Expanding explicitly the form of the matrix $\tilde{\Phi}$ in (\ref{eigenvalueequation2ndform}) and using the above fact, it follows that $E$ must satisfy the eigenvalue equation for the regularized Hamiltonian. Hence, we have a non-trivial solution of the above linear equation (\ref{eigenvalueequation2ndform}) for $\langle \tilde{f}_j(\epsilon)|\psi \rangle$ with some $|\psi\rangle \in L^2(\mathbb{R}^2)$ if and only if $\det \tilde{\Phi}(\epsilon, E)=\det \Phi(\epsilon, E) =0$.

\end{proof}

Now, we consider the limiting case as $\epsilon \rightarrow 0$ to properly  define the  initial formal Hamiltonian. For this reason, we choose $\lambda_1(\epsilon)$ in such a way that the regularized Hamiltonian has a reasonable and non-trivial limit as we remove the cut-off parameter, that is, as $\epsilon \to 0^+$.

\begin{myproposition} \label{proposition2} Let $\Phi_{ij}(E)=\lim_{\epsilon \to 0^+} \Phi_{ij}(\epsilon, E)$
and $E$ be a real negative number that satisfies $\det \Phi(E) \neq 0$. Then, the resolvent $R(\epsilon, E)$ of the regularized Hamiltonian $H_{a \Gamma, \, \epsilon}$ converges strongly to the expression $R(E)$, given by
\begin{eqnarray} \label{resolventcirclepoint}
 R(E):= R_0(E) + R_0(E) \sum_{i,j=1}^{2} | f_i \rangle \left(\Phi^{-1}(E) \right)_{ij} \langle f_j|  R_0(E) \;,
\end{eqnarray}
as $\epsilon \to 0^+$. Here for $j=1$, $\langle \mathbf{a} |R_0(E)$ is an integral operator whose kernel is given by  $R_0(\mathbf{a},\mathbf{x}|E)$ and for $j=2$, $\langle \boldsymbol{\Sigma} |R_0(E)$ is an integral operator whose kernel is given by $\frac{1}{L(\Gamma)} \int_{\Gamma} R_0(\mathbf{x},\boldsymbol{\gamma}(s)|E)\, d s $. 
\end{myproposition}

\begin{proof}

We begin the proof by calculating the matrix elements of $\Phi(\epsilon, E)$ in the limit $\epsilon \to 0^+$ for negative real values of $E$. The off-diagonal elements of the matrix $\Phi(\epsilon, E)$ for $E=-\nu^2$ with $\nu>0$ in the limit $\epsilon \rightarrow 0^+$ can be directly calculated using the Lebesgue dominated convergence theorem and the integral \cite{gradshteyn2014table}
\begin{eqnarray}
\int_{0}^{\infty} J_{\xi}(a x) J_{\xi}(b x) \frac{x}{x^2 + c^2} \; d x = \begin{cases} 
      K_{\xi}(a c) I_{\xi}(b c)   & 0<b<a \\
     K_{\xi}(b c) I_{\xi}(a c) & 0<a<b
      \end{cases} \;, \label{integralofJ0fraction}
\end{eqnarray}
for  $\Real(\xi)>-1$, so that
\begin{eqnarray}
\lim_{\epsilon \to 0^+} \Phi_{12}(\epsilon, -\nu^2) & = & \lim_{\epsilon \to 0^+} \Phi_{21}(\epsilon, -\nu^2)= - \lim_{\epsilon \to 0^+} \langle \mathbf{a}^{\epsilon}| R_0(-\nu^2)|\Gamma^{\epsilon} \rangle \nonumber \\ & = & - \frac{1}{2\pi} K_0\left(a \nu\right) I_0\left(R \nu \right) \;. 
\end{eqnarray}
The limit of the second diagonal term of the matrix $\Phi(\epsilon, -\nu^2)$ as $\epsilon \rightarrow 0^+$ can be evaluated easily thanks to the Lebesgue dominated convergence theorem so we have
\begin{eqnarray}
\lim_{\epsilon \to 0^+} \Phi_{22}(\epsilon, -\nu^2) & = & \frac{1}{\lambda_2}- \langle \Gamma^{\epsilon} |R_0(\epsilon, -\nu^2)|\Gamma^{\epsilon} \rangle = \frac{1}{\lambda_2}- \lim_{\epsilon \to 0^+} 
\int_{\mathbb{R}^2} \frac{|\langle \mathbf{p}|\Gamma^{\epsilon}\rangle |^2}{p^2 + \nu^2} \frac{d^2 p}{(2\pi)^2} \nonumber \\ & = & \frac{1}{\lambda_2}-- \frac{1}{2\pi} I_{0}(\nu R) K_0(\nu R) \;, \label{matrix22element}\end{eqnarray}
where we have used the result (\ref{pgammaepsilon}) and the continuity of the integral (\ref{integralofJ0fraction}) in the limiting case $a \to b$. The $\epsilon \to 0^+$ limit of the first diagonal element of the matrix $\Phi(\epsilon, E)$ in (\ref{regularizedPhi1}) includes the following term
\begin{eqnarray}
   \lim_{\epsilon \rightarrow 0} \langle \mathbf{a}^{\epsilon}|R_0(-\nu^2)|\mathbf{a}^{\epsilon} \rangle = \int_{0}^{\infty} K_t(\mathbf{a}, \mathbf{a}) e^{-t \nu^2} dt \;.
\end{eqnarray}
This is divergent due to the singular behaviour of the heat kernel around $t=0$ in two and three dimensions.
In line with the well-known idea of renormalization in field theory models, we  introduce a new parameter $\mu>0$ and make the following choice of the coupling constant $\lambda_1$ as a function of the regularization parameter $\epsilon$:   
\begin{eqnarray} \label{barecouplingconstant}
    \frac{1}{\lambda_1(\epsilon)}:= \int_{0}^{\infty} K_{t+\epsilon}(\mathbf{a},\mathbf{a}) e^{-t\mu^2} d t \;.
\end{eqnarray}
After substituting (\ref{barecouplingconstant}) for the  first diagonal element of the matrix (\ref{regularizedPhi1}) for $E=-\nu^2$ and then taking the limit as $\epsilon \rightarrow 0^+$, we get
\begin{eqnarray}
\lim_{\epsilon \to 0^+} \Phi_{11}(\epsilon, -\nu^2)= \frac{1}{4\pi} \log(\nu^2/\mu^2) \;.
\end{eqnarray}
Hence, we denote the limit of the matrix $\Phi(\epsilon, -\nu^2)$ as  $\Phi(-\nu^2)$, and it is given by
\begin{eqnarray}
 \Phi(-\nu^2) :=\left(
\begin{array}{cccc}
\frac{1}{4\pi} \log \left(\frac{\nu^2}{\mu^2}\right)& & -\frac{1}{2\pi} K_0\left(\nu a \right) I_0\left(\nu R \right) \\ \\
-\frac{1}{2\pi} K_0\left(\nu a \right) I_0\left(\nu R \right) & & \frac1{\lambda_2}- \frac{1}{2\pi} K_{0}(\nu R) I_0(\nu R) 
\end{array}
\right) \;. \label{formallimitofPhicirclept}
\end{eqnarray}
The formal limit of the resolvent (\ref{regularizedresolventcirclepoint}) of the regularized Hamiltonian as we take $\epsilon\to 0^+$,  is given by (\ref{resolventcirclepoint})
for $E =-\nu^2$, which satisfies $\det \Phi(E) \neq 0$ and the matrix $\Phi$ is given by (\ref{formallimitofPhicirclept}). Here $|f_1 \rangle = |\mathbf{a} \rangle$ and $|f_2 \rangle= | \Gamma \rangle$.

It remains to show that the above regularized resolvent (\ref{regularizedresolventcirclepoint}) converges strongly to the expression (\ref{resolventcirclepoint})
as $\epsilon \rightarrow 0^+$ for real negative values of $E$ that satisfy $\det \Phi(E)\neq 0$, that is,
\begin{eqnarray}
\lim_{\epsilon \to 0^+} || \left(R(\epsilon, E) - R(E)\right) |f \rangle|| = 0 \;,
\end{eqnarray}
for any $|f\rangle \in L^2(\mathbb{R}^2)$ and $E \in \rho(H_{a \Gamma, \, \epsilon})$. Since $E$ is assumed to satisfy $\det \Phi(E) \neq 0$, we conclude that $\det \Phi(\epsilon, E) \neq 0$ for sufficiently small $\epsilon>0$. Then, if we show that
\begin{eqnarray}
\lim_{\epsilon \to 0^+} ||R_0(E)|f_i(\epsilon)\rangle-R_0(E)|f_i \rangle||=0 \;,
\end{eqnarray}
 the strong convergence of the resolvent easily follows. The above condition is easy to check once we write it explicitly
\begin{eqnarray}
\lim_{\epsilon \to 0^+} ||R_0(E)|f_i(\epsilon)\rangle-R_0(E)|f_i \rangle||^2 = \begin{cases}
  \int_{\mathbb{R}^2} \left(1+e^{-\epsilon p^2} -2 e^{-\epsilon p^2/2} \right) \frac{1}{(p^2-E)^2}\frac{d^2 p}{(2\pi)^2} & \; \text{for} \; i=1 \\ \int_{\mathbb{R}^2} \left(1+e^{-\epsilon p^2} -2 e^{-\epsilon p^2/2} \right) \frac{J_{0}^{2}(p R)}{(p^2-E)^2}\frac{d^2 p}{(2\pi)^2} & \; \text{for} \; i=2  
\end{cases} \;,
\end{eqnarray}
where we have used the equations (\ref{paepsilon}) and (\ref{pgammaepsilon}). Then, the Lebesgue dominated convergence theorem implies that this limit is zero. 
\end{proof}

\begin{myremark}
Applying the method of renormalization to find well-defined results for  point Dirac delta potentials in two and three dimensions is actually well-known, see e.g. \cite{Huang, Jackiw}.  
\end{myremark}

\begin{myremark}
If the point delta term is absent in the model, the resulting resolvent associated with $H_0-\lambda_2 |\Gamma \rangle \langle \Gamma|$ is simply given by
\begin{eqnarray} \label{regularizedresolventcircle}
    R(E)= R_0(E) + R_0(E) | \Gamma \rangle \Phi^{-1}(E)  \langle \Gamma|  R_0(E) \;,
\end{eqnarray}
where the function $\Phi$ is given by equation (\ref{matrix22element}). Since the poles of the resolvent determine the bound state energies, zeroes of the function
\begin{eqnarray}
\Phi(E=-\nu^2)=\frac{1}{\lambda_2}-\frac{1}{2\pi} K_0(\nu R) I_0(\nu R) 
\end{eqnarray}
are the bound state energies of this system. This equation is exactly the same as the equation (2.4) given in \cite{exner2003spectra} written for the zero angular momentum sector up to rescaling of the coupling constant ($\alpha=\lambda_2/(2\pi R)$). In \cite{exner2003spectra}, all  angular momentum sectors are taken into account whereas the second interaction term that we consider here is the rank one perturbation  associated with the delta function supported on a circle. 
\end{myremark}

It is now natural to ask whether the above limiting expression is the resolvent of some self-adjoint operator.  This can be answered affirmatively by following the ideas given in \cite{Albeverio2012solvable} or in \cite{existence}. Here we essentially follow a similar argument presented in \cite{rajeevdimock} developed for the point interactions in the plane.

\begin{mylemma} \label{lemma3}
Let $E$ be a real negative number that satisfies $\det \Phi(E) \neq 0$. Then, $R(E)$ is invertible.
\end{mylemma}

\begin{proof}
We first show that the limit operator $R(E)$ for the real negative values of $E$ that satisfies $\det \Phi(E)\neq 0$ is invertible (equivalently, $\Ker(R(E))=\{|0 \rangle \}$). Suppose that $R(E)|f\rangle = |0 \rangle $ for some $|f \rangle \in L^2(\mathbb{R}^2)$. From the explicit expression of the operator $R(E)$ given by (\ref{resolventcirclepoint}) and writing the equation in momentum representation, we find
\begin{eqnarray} & & 
\hat{f}(\mathbf{p}) = - \sum_{i,j=1}^{2} \frac{\langle \mathbf{p}|f_i \rangle}{p^2-E} \left[\Phi^{-1}(E)\right]_{ij} 
\int_{\mathbb{R}^2} \langle f_j | R_0(E) | \mathbf{q} \rangle \langle \mathbf{q}|f \rangle \frac{d^2 q}{(2\pi)^2} \;.
\end{eqnarray}
By Cauchy-Schwarz inequality, we have
\begin{eqnarray} 
\int_{\mathbb{R}^2} \langle \mathbf{a} | R_0(E) | \mathbf{q} \rangle \langle \mathbf{q}|\psi \rangle \frac{d^2 q}{(2\pi)^2} & = &
\int_{\mathbb{R}^2} \frac{e^{i \mathbf{q} \cdot \mathbf{a}}}{q^2-E} f(\mathbf{q}) \frac{d^2 q}{(2\pi)^2} \nonumber \\ & \leq & \left(\int_{0}^{\infty} \frac{q}{(q^2-E)^2}\; \frac{dq}{2\pi}\right)^{1/2} ||f|| < \infty \;, \label{csbound1}
\end{eqnarray}
and 
\begin{eqnarray} 
\int_{\mathbb{R}^2} \langle \Gamma | R_0(E) | \mathbf{q} \rangle \langle \mathbf{q}|\psi \rangle \frac{d^2 q}{(2\pi)^2} & = &
\int_{\mathbb{R}^2} \frac{J_{0}(q R)}{q^2-E} f(\mathbf{q}) \frac{d^2 q}{(2\pi)^2} \nonumber \\ & \leq & \left(\int_{0}^{\infty} \frac{q}{(q^2-E)^2}\; \frac{dq}{2\pi}\right)^{1/2} ||f|| < \infty \;. \label{csbound2}
\end{eqnarray}
With the above bounds (\ref{csbound1}) and (\ref{csbound2}), we show that  
\begin{eqnarray} & & \hskip-2cm
\hat{f}(\mathbf{p}) = - \Bigg[ e^{-i \mathbf{p} \cdot \mathbf{a}} \left[\Phi^{-1}(E)\right]_{11} C_1 + e^{-i \mathbf{p} \cdot \mathbf{a}} \left[\Phi^{-1}(E)\right]_{12} C_2 \nonumber \\ & & \hspace{3cm} + \, J_{0}(p R) \left[\Phi^{-1}(E)\right]_{21} C_2 + J_{0}(p R) \left[\Phi^{-1}(E)\right]_{22} C_1  \Bigg] \;,
\end{eqnarray}
where $C_1$, $C_2$ are finite real numbers and $E$ is a negative real number that satisfies $\det \Phi(E) \neq 0$. However, this solution $\hat{f}(\mathbf{p})$ can not be in $L^2(\mathbb{R}^2)$ unless $|f \rangle = |0 \rangle$. %
\end{proof}

This Lemma allows us to define an operator $H$ depending on the parameter $\mu$ via 
\begin{eqnarray}
R(E):= (H_{a \Gamma}(\mu)-E)^{-1}
\end{eqnarray}
for the above values of $E$. From now on we suppress the dependence of the Hamiltonian on the parameter $\mu$ for simplicity.

\begin{mytheo} \label{theo1} For complex $E$ not in $\det \Phi(E)=0$ and $[0, \infty)$, the resolvent $R(\epsilon, E)$ of regularized Hamiltonian $H_{a \Gamma, \, \epsilon}$ converges strongly to $R(E)$. Furthermore, there exists a self-adjoint operator $H_{a \Gamma}$ such that $R(E)=(H_{a \Gamma}-E)^{-1}$. 
\end{mytheo}

\begin{proof}

Using the above preliminary steps together with a version of Trotter-Kato theorem, quoted in Appendix (see also \cite{rajeevdimock}), it follows that the limit $R(\epsilon, E)$ converges strongly to $R(E)$ as $\epsilon \to 0^+$ for all complex numbers $E$ except for the interval $[0, \infty)$ and for the values satisfying $\det \Phi(E)\neq 0$. Moreover, there exists a self-adjoint operator $H_{a \Gamma}$ such that $R(E)=(H_{a \Gamma}-E)^{-1}$ and the matrix $\Phi$ for complex values are defined through its analytic continuation, given by
\begin{eqnarray}
 \Phi(k^2)=\left(
\begin{array}{cccc}
\frac{1}{4\pi} \log \left(-\frac{k^2}{\mu^2}\right)& & -\frac{1}{2\pi} K_0\left(-i k a \right) I_0\left(-i k R \right) \\ \\
-\frac{1}{2\pi} K_0\left(-i k a \right) I_0\left(-i k R\right) & & \frac1{\lambda_2}- \frac{1}{2\pi} K_{0}(- i k R) I_0(- i k R)
\end{array}
\right) \;, \label{principalmatrixpointcircle}
\end{eqnarray}
where we parametrize $E=k^2$ with unambiguous square root $k$ with $\Imaginary(k)>0$ for convenience. We shall call this matrix as principal matrix from now on. 

\end{proof}

\begin{mytheo} \label{theo2} The domain of the self-adjoint operator $H_{a \Gamma}$ defined by its resolvent $R(E)=(H_{a \Gamma}-E)^{-1}$ consists of all functions $\psi(\mathbf{r})$ in the following form for $\mathbf{r} \in \mathbb{R}^2 \setminus \{\mathbf{a} \cup \Gamma\}$
\begin{eqnarray}
\psi(\mathbf{r}) = \phi_k(\mathbf{r}) +\sum_{i,j=1}^{2} F_i(\mathbf{r}) \left[\Phi^{-1}(k^2)\right]_{ij} \langle f_j|\phi_k \rangle \;,
\end{eqnarray}
where $F_i(\mathbf{r})=\langle \mathbf{r} | R_0(k^2) |f_i \rangle$, given explicitly by
\begin{eqnarray} \hskip-1cm F_1(\mathbf{r})= 
\langle \mathbf{r}|R_0(k^2)|f_1 \rangle = \langle \mathbf{r}|R_0(k^2)|\mathbf{a}\rangle  & =&  \int_{\mathbb{R}^2} \frac{e^{i \mathbf{p} \cdot (\mathbf{r}-\mathbf{a})}}{p^2-k^2} \; \frac{d^2 p}{(2\pi)^2} = \frac{i}{4} H_{0}^{(1)}(k|\mathbf{r}-\mathbf{a}|) \;, \label{rR0f1} \\ 
F_2(\mathbf{r}) = \langle \mathbf{r}|R_0(k^2)|f_2 \rangle =
\langle \mathbf{r}|R_0(k^2)|\Gamma \rangle & = & \int_{\mathbb{R}^2} \frac{e^{i \mathbf{p} \cdot \mathbf{r}}}{p^2-k^2} \;J_0(p R) \frac{d^2 p}{(2\pi)^2} \nonumber \\ & = & \int_{0}^{\infty} \frac{p J_0(p r) J_0(p R)}{p^2-k^2} \frac{d p}{(2\pi)} \nonumber \\ & & \hskip-2cm =  \frac{i}{4} \left(H_{0}^{(1)}(k r) J_0(k R) \theta(R-r) +H_{0}^{(1)}(k R) J_0(k r) \theta(r-R)  \right) \;. \label{rR0f2}
\end{eqnarray}
Here $\phi_k \in D(H_0)=H^{2}(\mathbb{R}^2)$ and $k^2 \in \rho(H)$ with $\Imaginary(k)>0$. The above decomposition is unique and 
$(H_{a \Gamma}-k^2)|\psi\rangle = (H_0-k^2)|\phi_k\rangle$. Moreover, suppose that $D(H_{a \Gamma}) \ni \psi(\mathbf{r})=0$ in an open set $U \subseteq \mathbb{R}^2$. Then, $H_{a \Gamma}\psi(\mathbf{r})=0$ for all $\mathbf{r} \in U$.
\end{mytheo}

\begin{proof}

Suppose $E=k^2$ with unambiguous square root $k$ where $\Imaginary(k)>0$ and $\phi_k(\mathbf{r}) \in D(H_0)=H^{2}(\mathbb{R}^2)$.
Thanks to the self-adjointness of $H$, we have
\begin{eqnarray}
D(H_{a \Gamma}) = (H_{a \Gamma}-k^2)^{-1} L^2(\mathbb{R}^2) = (H_{a \Gamma}-k^2)^{-1}(H_0-k^2)D(H_0) \;.
\end{eqnarray}
Then, using the explicit form of the resolvent formula (\ref{resolventcirclepoint}), we have the following characterization of the domain of $H_{a \Gamma}$:
\begin{eqnarray}
D(H_{a \Gamma})= \left( 1+ \sum_{i,j=1}^{2} R_0(k^2) | f_i \rangle \left[\Phi^{-1}(k^2)\right]_{ij} \langle f_j| \right) D(H_0) \;.
\end{eqnarray}
This means that the domain of $H_{a \Gamma}$ consists of all functions of the following form
\begin{eqnarray}
\psi(\mathbf{r}) = \phi_k(\mathbf{r}) +\sum_{i,j=1}^{2} \langle \mathbf{r} | R_0(k^2) |f_i \rangle \left[\Phi^{-1}(k^2)\right]_{ij} \langle f_j|\phi_k \rangle \;,
\end{eqnarray}
where $\langle \mathbf{a}|\phi_k\rangle=\phi_k(\mathbf{a})$, $\langle \Gamma|\phi_k \rangle = \frac{1}{L} \int_{S^1} \phi_k(\boldsymbol{\gamma}(s)) \; d s$. Note that the point evaluation and integral over the curve here  for $\phi_k \in H^2(\mathbb{R}^2)$ are well defined thanks to the Sobolev embedding theorem $H^2(\mathbb{R}^n) \hookrightarrow C^l(\mathbb{R}^n)$ with the condition $2>l+\frac{n}{2}$ \cite{Evans, haroske2007distributions}, that is, there exist unique continuous representatives for (equivalence classes of) functions in $H^2(\mathbb{R}^n)$ and we use this representative for $\phi_k$ here.  
We have evaluated the last integral by the analytic continuation of the result (\ref{integralofJ0fraction}) and used the fact that $K_0(z)= \frac{i \pi}{2} H_{0}^{(1)}(e^{i \pi/2} z)$ and $I_0(z)=e^{-i \pi/2} J_0(e^{i \pi/2}z)$ for $-\pi < arg(z) < \pi/2$ \cite{Lebedev1965special}, where $H_{0}^{(1)}$ is the zeroth order Hankel function of the first kind. Hence, we obtain
\begin{eqnarray} & & \hskip-1cm
\psi(\mathbf{r}) = \phi_k(\mathbf{r}) + \frac{i}{4} H_{0}^{(1)}(k|\mathbf{r}-\mathbf{a}|) \Bigg(\left[\Phi^{-1}(k^2)\right]_{11} \phi_k(\mathbf{a}) + \left[\Phi^{-1}(k^2)\right]_{12} \bigg(\frac{1}{L} \int_{S^1} \phi_k(\boldsymbol{\gamma}(s))\; d s \bigg)  \Bigg) \nonumber \\ & &  \hspace{2cm} + \, \frac{i}{4} \left(H_{0}^{(1)}(k r) J_0(k R) \theta(R-r) +H_{0}^{(1)}(k R) J_0(k r) \theta(r-R)  \right) \nonumber \\ & & \hspace{3cm} \times \, \Bigg(\left[\Phi^{-1}(k^2)\right]_{21} \phi_k(\mathbf{a}) + \left[\Phi^{-1}(k^2)\right]_{22} \bigg(\frac{1}{L} \int_{S^1} \phi_k(\boldsymbol{\gamma}(s))\; d s \bigg)  \Bigg) \;. \label{domaindecomposition}
\end{eqnarray}
Indeed, this decomposition (\ref{domaindecomposition}) is unique. For this, let us set $\psi(\mathbf{r})=0$ identically. Then, it follows from the above decomposition that 
\begin{eqnarray} & & \hskip-1cm
\phi_k(\mathbf{r})= - \frac{i}{4} H_{0}^{(1)}(k|\mathbf{r}-\mathbf{a}|) \Bigg(\left[\Phi^{-1}(k^2)\right]_{11} \phi_k(\mathbf{a}) + \left[\Phi^{-1}(k^2)\right]_{12} \bigg(\frac{1}{L} \int_{S^1} \phi_k(\boldsymbol{\gamma}(s))\; d s \bigg)  \Bigg) \nonumber \\ & &  \hspace{2cm} - \, \frac{i}{4} \left(H_{0}^{(1)}(k r) J_0(k R) \theta(R-r) +H_{0}^{(1)}(k R) J_0(k r) \theta(r-R)  \right) \nonumber \\ & & \hspace{3cm} \times \, \Bigg(\left[\Phi^{-1}(k^2)\right]_{21} \phi_k(\mathbf{a}) + \left[\Phi^{-1}(k^2)\right]_{22} \bigg(\frac{1}{L} \int_{S^1} \phi_k(\boldsymbol{\gamma}(s))\; d s \bigg)  \Bigg) \;. \label{domaindecomposition2}
\end{eqnarray}
Since the functions $H_{0}^{(1)}(k|\mathbf{r}-\mathbf{a}|)$ and $H_{0}^{(1)}(k r) J_0(k R) \theta(R-r) +H_{0}^{(1)}(k R) J_0(k r) \theta(r-R)$  in each term are discontinuous at $\mathbf{r}=\mathbf{a}$ and $r=R$, the function $\phi_k(\mathbf{r})$ can only be continuous if 
\begin{eqnarray}
\left[\Phi^{-1}(k^2)\right]_{11} \phi_k(\mathbf{a}) + \left[\Phi^{-1}(k^2)\right]_{12} \bigg(\frac{1}{L} \int_{S^1} \phi_k(\boldsymbol{\gamma}(s))\; d s \bigg)  & = & 0 \;, \\ \left[\Phi^{-1}(k^2)\right]_{21} \phi_k(\mathbf{a}) + \left[\Phi^{-1}(k^2)\right]_{22} \bigg(\frac{1}{L} \int_{S^1} \phi_k(\boldsymbol{\gamma}(s))\; d s \bigg)  & = & 0 \;. 
\end{eqnarray}
Therefore, these conditions imply that the decomposition (\ref{domaindecomposition}) is unique. It is also straightforward to show that $(H_{a \Gamma}-k^2)^{-1}(H_0-k^2)|\phi_k \rangle =|\psi \rangle$, which is equivalent to $(H_{a \Gamma}-k^2)|\psi \rangle = (H_0-k^2)|\phi_k \rangle$.

After showing the existence of a self-adjoint operator $H_{a \Gamma}$ associated with the resolvent $R(E)$, we may not guarantee that $H_{a \Gamma}$ must be of the form $H_0+V$ with some operator $V$. Nevertheless, we can show that $H_{a \Gamma}$ is a local operator in the sense that $\psi(\mathbf{r})=0$ in an open set $U \subseteq \mathbb{R}^2$ implies that $H_{a \Gamma}\psi(\mathbf{r})= \langle \mathbf{r}|H_{a \Gamma}|\psi\rangle = 0$. For this, let $\psi(\mathbf{r})=0$ for all $\mathbf{r} \in U$. Then, the function $\phi_k(\mathbf{r})$ for $\mathbf{r} \in U$ is given by equation (\ref{domaindecomposition2}). If $U \cap \{\mathbf{a} \cup \Gamma \} = \emptyset$, the action of $H_0-k^2$ onto the function $\phi_k(\mathbf{r})$ vanishes. Since 
$H_{0}^{(1)}$ is the Green's function of Helmholtz equation in two dimensions and $J_0(k r)$ satisfies Helmholtz equation 
we get $H_{a \Gamma}\psi(\mathbf{r})= k^2 \psi(\mathbf{r})+(H_0-k^2)\phi_k(\mathbf{r})=0$ in $U$. For the case $\mathbf{a} \in U$, the continuity of the function $\phi_k$ at $\mathbf{r}=\mathbf{a}$ from the equation (\ref{domaindecomposition2}) implies that $\left[\Phi^{-1}(k^2)\right]_{11} \phi_k(\mathbf{a}) + \left[\Phi^{-1}(k^2)\right]_{12} \bigg(\frac{1}{L} \int_{S^1} \phi_k(\boldsymbol{\gamma}(s))\; d s \bigg)=0$. Similarly, if $\Gamma \in U$, the term $\left[\Phi^{-1}(k^2)\right]_{21} \phi_k(\mathbf{a}) + \left[\Phi^{-1}(k^2)\right]_{22} \bigg(\frac{1}{L} \int_{S^1} \phi_k(\boldsymbol{\gamma}(s))\; d s \bigg)$ must vanish. Hence, we obtain $H_{a \Gamma} \psi(\mathbf{r})=0$ in $U$.

\end{proof}

\subsection{Bound State Analysis}
\label{Bound State Analysis for Circle and Point}

\begin{mytheo} Let $\mathbf{a} \in \mathbb{R}^2$ and $\Gamma$ be the circle centered at the origin with radius $R<a$. Then, the essential spectrum of $H_{a \Gamma}$ associated with the point delta and the rank one perturbation supported by $\Gamma$ coincides with the essential spectrum of the free Hamiltonian, i.e., $\sigma_{ess}(H_{a \Gamma})=\sigma_{ess}(H_0)=[0, \infty)$. Furthermore, the point spectrum $\sigma_p(H_{a \Gamma})$ of $H_{a \Gamma}$ lies in the negative real axis and $H_{a \Gamma}$ has at least one and at most two negative eigenvalues (counting multiplicity). Let $\Real(k)=0$ and $\Imaginary(k)>0$, then $k^2 \in \sigma_p(H_{a \Gamma})$ if and only if $\det \Phi(k^2)=0$ and multiplicity (degeneracy) of the eigenvalue $k^2$ is the same as the multiplicity of this zero eigenvalue of the matrix $\Phi(k^2)$. Moreover, let $E=-\nu_{*}^2<0$ be an eigenvalue of $H_{a \Gamma}$, then the eigenfunction $|\psi_{ev}\rangle$ associated with this eigenvalue is given by
\begin{eqnarray*}
\psi_{ev}(\mathbf{r})= \sum_{i=1}^{2} \langle \mathbf{r}| R_0(-\nu_{*}^2)|f_i \rangle A_i \;,
\end{eqnarray*}
where $(A_1, A_2)$ is an  eigenvector corresponding to a zero eigenvalue of $\Phi(-\nu_{*}^2)$.
\end{mytheo}

\begin{proof}

As it is well-known that the point spectrum $\sigma_p$ of an operator $H_{a \Gamma}$ consists of the set of complex numbers $E$ such that $\Ker(H_{a \Gamma}-E)\neq \{|0 \rangle\}$. From the explicit expression of the resolvent $R(k^2)$ given by (\ref{resolventcirclepoint}) for $E=k^2$, the poles of the resolvent for $k^2<0$ can only appear if the matrix $\Phi(k^2)$ is singular, that is, if 
\begin{eqnarray}
\det \Phi(k^2)=0 \;.\label{boundstatecondition}
\end{eqnarray}
Let $|\psi_{ev} \rangle$ be an eigenvector of $H_{a \Gamma}$ with corresponding eigenvalue $E_{ev}=k_{ev}^{2}$, i.e.,
\begin{eqnarray}
H_{a \Gamma}|\psi_{ev} \rangle = E_{ev} |\psi_{ev} \rangle \;,
\end{eqnarray}
where $|\psi_{ev} \rangle \in D(H_{a \Gamma})$. Since any function in the domain of $H_{a \Gamma}$ can be decomposed according to Theorem \ref{theo2}, we have
\begin{eqnarray}
|\psi_{ev}\rangle = |\phi_k \rangle + \sum_{i,j=1}^{2}  R_0(k^2) |f_i \rangle \left[\Phi^{-1}(k^2)\right]_{ij} \langle f_j|\phi_k \rangle \;,
\label{psiev} \end{eqnarray}
for some $k^2 \in \rho(H_{a \Gamma})$ with $\Imaginary(k)>0$ and $|\phi_k\rangle \in D(H_0)$. Actually, Theorem \ref{theo2} provides us  another relation between $|\psi_{ev}\rangle$ and $|\phi_k\rangle$:
\begin{eqnarray}
|\phi_k\rangle = (k_{ev}^{2}- k^2)R_0(k^2)|\psi_{ev}\rangle \;. \label{psievphik}
\end{eqnarray}
Substituting equation (\ref{psiev}) into (\ref{psievphik}), we find
\begin{eqnarray}
|\phi_k \rangle = (k_{ev}^{2}-k^2) \Bigg(R_0(k^2) |\phi_k \rangle + \sum_{i,j=1}^{2} R_0(k^2) R_0(k^2) |f_i \rangle \left[\Phi^{-1}(k^2)\right]_{ij} \langle f_j|\phi_k \rangle \Bigg) \;. \label{psievphik2}
\end{eqnarray}
By acting $H_0-k^2$ on this vector, it yields
\begin{eqnarray}
(H_0-k_{ev}^{2})|\phi_k \rangle = (k_{ev}^{2}-k^2) \sum_{i,j=1}^{2} R_0(k^2) |f_i \rangle \left[\Phi^{-1}(k^2)\right]_{ij} \langle f_j|\phi_k \rangle \;. \label{solutionphik}
\end{eqnarray}
The solution of this in momentum representation is given by
\begin{eqnarray}
\hat{\phi}_k(\mathbf{p})= \frac{(k_{ev}^{2}-k^2)}{p^2-k_{ev}^{2}} \sum_{i,j=1}^{2}  \frac{\langle \mathbf{p}|f_i \rangle}{p^2-k^2} \left[\Phi^{-1}(k^2)\right]_{ij} \langle f_j|\phi_k \rangle \;. \label{solutionphikmomentumspace}
\end{eqnarray}
If $E_{ev}=k_{ev}^2 \geq 0$, then this equation has no nontrivial solution since $\hat{\phi}_k(\mathbf{p})$ cannot lie in $L^2(\mathbb{R}^2)$ unless it is identically zero. This implies that $|\phi_k \rangle \notin L^2(\mathbb{R}^2)$ thanks to the Plancherel theorem. Hence, $\psi_{ev}(\mathbf{r})=0$ for all $\mathbf{r} \in \mathbb{R}^2$, which proves that there is no nonnegative eigenvalue of $H_{a \Gamma}$.

However, if $E_{ev}=k_{ev}^{2}=-\nu_{*}^2<0$ with $\nu>0$, it is legitimate to apply $R_0(-\nu_{*}^2)$ on each side of equation (\ref{solutionphik}) and get
\begin{eqnarray}
|\phi_k \rangle = \left(R_0(-\nu_{*}^2) - R_0(k^2)\right) \sum_{i,j=1}^{2}  |f_i \rangle \left[\Phi^{-1}(k^2)\right]_{ij} \langle f_j|\phi_k \rangle \;. \label{phiksolutionbound}
\end{eqnarray}
Inserting this solution into (\ref{psiev}), we formally find the eigenfunctions of $H_{a \Gamma}$
\begin{eqnarray}
\psi_{ev}(\mathbf{r})= \sum_{i,j=1}^{2} \langle \mathbf{r} | R_0(-\nu_{*}^2)|f_i \rangle \left[\Phi^{-1}(k^2)\right]_{ij} \langle f_j|\phi_k \rangle \;. \label{eigenfunctionsolution}
\end{eqnarray}
This formal solution includes unknown factors $\langle f_j|\phi_k \rangle$. In order to find them, we first note that the principal matrix $\Phi$ can also be expressed purely in terms of the free resolvent kernels, that is,
\begin{eqnarray} \label{Phifreeresolventkernel}
\Phi(k^2) = \left(
\begin{array}{cccc}
\langle f_1|\left( R_0(-\mu^2)-R_0(k^2)\right) |f_1 \rangle & & - \langle f_1| R_0(k^2) |f_2 \rangle \\ \\
- \langle f_2| R_0(k^2) |f_1 \rangle & & \frac1{\lambda_2}-\langle f_2|R_0(k^2) |f_2 \rangle 
\end{array}
\right) \;. 
\end{eqnarray}
Then, it is easy to check that 
\begin{eqnarray}
\langle f_i|\left( R_0(-\nu_{*}^2)-R_0(k^2)\right) |f_j \rangle = \Phi_{ij}(k^2)-\Phi_{ij}(-\nu_{*}^2) \;. \label{differenceinresolventkernel}
\end{eqnarray}
Using this result in (\ref{phiksolutionbound}) after taking the projection onto $\langle f_{j'}|$, we obtain
\begin{eqnarray}
\sum_{j=1}^{2} \Phi_{ij}(-\nu_{*}^2) A_j = 0 \;, \label{zeroeigenvalueofPhi}
\end{eqnarray}
where $A_j= \sum_{i=1}^{2} \left[\Phi^{-1}(k^2)\right]_{ji}\langle f_i |\phi_k \rangle$. This equation tells us that $A_i$ is an eigenvector of the matrix $\Phi(-\nu_{*}^2)$ with a zero eigenvalue.

Conversely, let us suppose that
\begin{eqnarray}
|\psi_{ev} \rangle =\sum_{i=1}^{2} R_0(-\nu_{*}^2)|f_i \rangle A_i \;, \label{psievnew}
\end{eqnarray}
where $A_i=\sum_{j=1}^{2} \left[\Phi^{-1}(k^2)\right]_{ij}\langle f_j |\phi_k \rangle$ is an eigenvector of $\Phi(-\nu_{*}^2)$ with eigenvalue zero. We will show that $|\psi_{ev} \rangle \in D(H_{a \Gamma})$ and $H_{a \Gamma}|\psi_{ev}\rangle = -\nu_{*}^2 |\psi_{ev} \rangle$. First, we need to show that $|\psi_{ev}\rangle \in D(H_{a \Gamma})$. For this, we define
\begin{eqnarray}
|\phi_k \rangle = (k_{ev}^{2}-k^2)R_0(k^2)|\psi_{ev}\rangle  \label{phiknew}
\end{eqnarray}
for some $k^2 \in \rho(H_{a \Gamma})$ with $\Imaginary(k)>0$. Then, it follows easily that $|\phi_k \rangle \in D(H_0)$ and inserting (\ref{psievnew}) into (\ref{phiknew}) and using the first resolvent identity for the free resolvent we obtain 
\begin{eqnarray}
|\phi_k \rangle = \left(R_0(-\nu_{*}^2)-R_0(k^2)\right) \sum_{i=1}^{2} |f_i \rangle A_i \;\label{phiknew2}
\end{eqnarray}
or
\begin{eqnarray}
|\phi_k \rangle + R_0(k^2) \sum_{i=1}^{2} |f_i \rangle A_i = |\psi_{ev}\rangle \;.
\end{eqnarray}
Moreover, by taking the projection of (\ref{phiknew}) onto $\langle f_{j}|$ and using the above result (\ref{differenceinresolventkernel}), $|\psi_{ev} \rangle \in D(H_{a \Gamma})$ by Theorem \ref{theo2}. Finally,  using the result $(H_0-k^2)|\phi_k \rangle = (H_{a \Gamma}-k^2)|\psi_{ev} \rangle$ in Theorem \ref{theo2} for the eigenstate $|\psi_{ev}\rangle$, and equation (\ref{phiknew}) we deduce that
\begin{eqnarray}
H_{a \Gamma}|\psi_{ev}\rangle = (H_0-k^2)|\phi_k \rangle + k^2 |\psi_{ev}\rangle = -\nu_{*}^2 |\psi_{ev}\rangle \;.
\end{eqnarray}

It is useful to express the condition (\ref{boundstatecondition}) in terms of a real positive parameter $\nu$, defined by $\nu=-i k>0$. Then, the solutions of the equation $\det \Phi(-\nu^2)=0$ determine the point spectrum of $H_{a \Gamma}$ or bound state spectrum of $H_{a \Gamma}$. 
However, finding the roots of equation (\ref{boundstatecondition}) is analytically not possible. Nevertheless, we may obtain some information about the bound states as follows. First, suppose that the principal matrix $\Phi$ has an eigenvector $A$ associated with the eigenvalue $\omega$, 
\begin{eqnarray}
\Phi A =\omega A \;. 
\end{eqnarray}
The eigenvalues can be explicitly calculated
\begin{align}
    \omega_{1}(\nu) & = \frac{1}{4 \pi  \lambda } \Bigg\{ 2 \pi + \lambda  \log \left(\frac{\nu }{\mu }\right)-\lambda  I_0(\nu R) K_0(\nu R) - \Bigg[ \lambda^2 I_{0}^{2}(\nu R) \left(4 K_{0}^{2}(\nu a)+ K_{0}^{2}(\nu R)\right) \nonumber \\ & \hspace{2cm} + \left(\lambda  \log \left(\frac{\nu }{\mu }\right)-2 \pi \right)^2      + 2 \lambda  I_0(\nu R) K_0(\nu R) \left(\lambda  \log \left(\frac{\nu }{\mu }\right)-2 \pi \right)\Bigg]^{1/2} \Bigg\} \;,
\end{align}
and 
\begin{align}
    \omega_{2}(\nu) & = \frac{1}{4 \pi  \lambda } \Bigg\{ 2 \pi + \lambda  \log \left(\frac{\nu }{\mu }\right)-\lambda  I_0(\nu R) K_0(\nu R) + \Bigg[ \lambda^2 I_{0}^{2}(\nu R) \left(4 K_{0}^{2}(\nu a)+ K_{0}^{2}(\nu R)\right) \nonumber \\ & \hspace{2cm} + \left(\lambda  \log \left(\frac{\nu }{\mu }\right)-2 \pi \right)^2      + 2 \lambda  I_0(\nu R) K_0(\nu R) \left(\lambda  \log \left(\frac{\nu }{\mu }\right)-2 \pi \right)\Bigg]^{1/2} \Bigg\} \;.
\end{align}
Finding zeroes of the determinant of the matrix $\Phi$ is equivalent to finding the zeroes of its eigenvalues. We now show that these are increasing functions of $\nu$ by expressing the principal matrix $\Phi$ in its closed form. Suppose for simplicity that the eigenvectors $A$ are normalized. We can determine how the eigenvalues change with respect to $\nu$ according to the Feynman-Hellman theorem \cite{thirring2013quantum}
\begin{equation} \label{derivativeofeigenvalue}
    \frac{\partial \omega}{\partial \nu} =  A^{*T} \frac{\partial \Phi}{\partial \nu} A \;,
\end{equation}
where $*$ and $T$ denote the complex conjugation and transpose, respectively. Here, it is convenient to express the derivative of the principal matrix in the following form,  
\begin{eqnarray}
\frac{\partial \Phi_{11}}{\partial \nu} & = & \frac{1}{2\pi \nu} \;, \\
\frac{\partial \Phi_{12}}{\partial \nu} = \frac{\partial \Phi_{21}^{*}}{\partial \nu} & = & (2\nu) \int_{\mathbb{R}^2} \frac{e^{i \mathbf{p} \cdot \mathbf{a}}}{(p^2 + \nu^2)^2} J_{0}(p R) \frac{d^2 p}{(2\pi)^2} \;, \\  \frac{\partial \Phi_{22}}{\partial \nu} & = &
 (2\nu) \int_{\mathbb{R}^2} \frac{J_{0}^{2}(p R)}{(p^2 + \nu^2)^2}  \frac{d^2 p}{(2\pi)^2}   \;, \end{eqnarray}
by taking the derivative of $\Phi$ under the integral sign thanks to the Lebesgue dominated convergence theorem. Then, one can show that 
\begin{eqnarray}
\frac{\partial \omega}{\partial \nu} = (2\nu) \int_{\mathbb{R}^2} \bigg| A_1 e^{i \mathbf{p} \cdot \mathbf{a}} + A_2 J_0(p R) \bigg|^2 \; \frac{1}{(p^2 + \nu^2)^2} \; \frac{d^2 p}{(2\pi)^2} > 0 \;, \label{eigenvaluesincreasing}
\end{eqnarray}
for all $\nu>0$, that is, all the eigenvalues $\omega$ of the principal matrix $\Phi$ are strictly increasing functions of $\nu$.

The solutions of (\ref{boundstatecondition}) can also be considered as the zeroes of the eigenvalues of the principal matrix $\Phi$ so all the bound state energies can be found from the zeroes of the eigenvalues, say $\nu_*$, for which  
\begin{eqnarray}
E=-\nu_{*}^{2} \;.
\end{eqnarray}
The positivity condition (\ref{eigenvaluesincreasing}) implies that there are at most two bound state energies since each eigenvalue can cross the $\nu$ axis  only once. The zero of the eigenvalue $\omega_1$ corresponds to the ground state energy.
This bound state always exists for all values of the parameter since $\lim_{\nu \rightarrow 0^+} \omega_1=-\infty$ and it is an increasing function of $\nu$ and positive for sufficiently large values of $\nu$. However, the second eigenvalue $\omega_2$ may not have any zeroes if it is not negative around $\nu=0$.

It follows from Weyl's theorem \cite{reedsimonv4} that the essential spectra of $H_{a \Gamma}$ and $H_0$ coincide, that is, $\sigma_{ess}(H_{a \Gamma})=\sigma_{ess}(H_0)=[0,\infty)$ if we show that $R(E)-R_0(E)$ is compact for some $E \in \rho(H_{a \Gamma}) \cap \rho(H_0)$.
Note that we have $R(E)-R_0(E)$ given by an explicit formula (\ref{resolventcirclepoint}). Here $\Phi_{ij}$  is invertible for a sufficiently negative $E_*$ on the real axis and all its eigenvalues become positive. 
Therefore 
\begin{equation}
 R(E_*)-R_0(E_*)=   R_0(E_*) \sum_{i,j=1}^{2} |f_i\rangle \Phi^{-1}_{ij} \langle f_j| R_0(E_*) \;
\end{equation}
indeed becomes a finite rank operator. For this, note that the principal matrix has a spectral decomposition $\Phi^{-1}(E_*)=\sum_k \omega_k^{-1}(E_*) A^{(k)}(E_*)A^{(k)*}(E_*)$ with $A^{(k)}$ representing the $k$th eigenvector of $\Phi(E)$ and $\omega_k$ is the corresponding eigenvalue. All the eigenvalues become positive for $E_*$. We therefore need to observe that all the vectors 
\begin{eqnarray}
\sum_{i=1}^{2} \omega_k^{-1/2}(E_*) A_i^{(k)}R_0(E_*)|f_i\rangle \;
\end{eqnarray}
for $k=1,2$ have a finite norm, as can be seen as follows;
\begin{eqnarray}
|| \sum_{i=1}^{2} \omega_k^{-1/2}(E_*) A^{(k)}_iR_0(E_*)|f_i\rangle||\leq \sum_{i=1}^{2} |\omega_k^{-1/2}(E_*) A_i^{(k)}|||R_0(E_*)|f_i\rangle|| \;.
\end{eqnarray}
Hence, we have shown that $R(E)-R_0(E)$ is a trace class operator, which is compact. 

\end{proof}

\begin{myremark}
Figure \ref{fig:eigenvaluesptcircle} below shows how the eigenvalues change with respect to $\nu$ for the particular values of parameters. 
\begin{figure}[h!]
    \centering
    \includegraphics{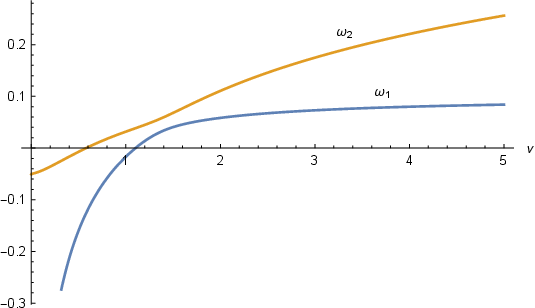}
    \caption{Eigenvalues of the principal matrix $\Phi$ versus $\nu$ for $\lambda=10, \mu=1, R=1, a=2$ units.}
    \label{fig:eigenvaluesptcircle}
\end{figure}
One can also numerically calculate the bound state energies and plot them as a function of $a$ and $R$, respectively for the fixed given values of the parameters, as shown in Figs. \ref{fig:ptcircleenergyvsa} and \ref{fig:ptcircleenergyvsR}.
\begin{figure}[h!]
     \centering
     \begin{subfigure}[b]{0.45\textwidth}
         \centering
         \includegraphics[width=\textwidth]{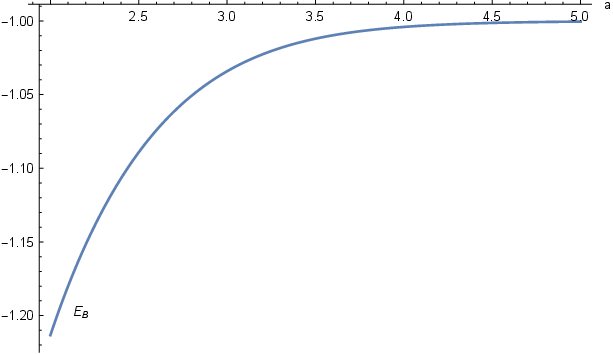}
         \caption{Ground state energy versus $a$}
         \label{fig:energyvsa1}
     \end{subfigure}
     \hfill
     \begin{subfigure}[b]{0.45\textwidth}
         \centering
         \includegraphics[width=\textwidth]{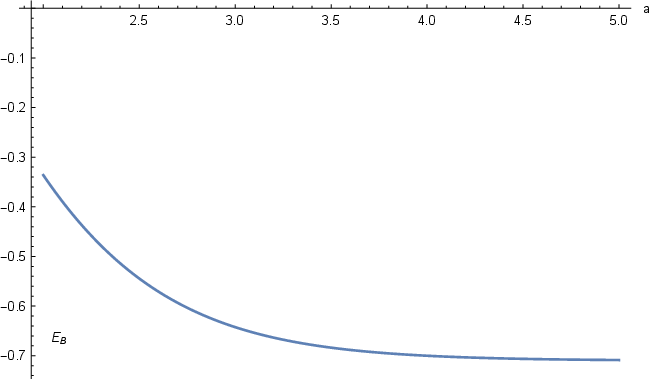}
         \caption{Excited state energy versus $a$}
         \label{fig:energyvsa2}
     \end{subfigure}
        \caption{Bound state energies versus $a$ for  $\lambda=10, R=1, \mu=1$ units.}
        \label{fig:ptcircleenergyvsa}
\end{figure}
\begin{figure}[h!]
     \centering
     \begin{subfigure}[b]{0.45\textwidth}
         \centering
         \includegraphics[width=\textwidth]{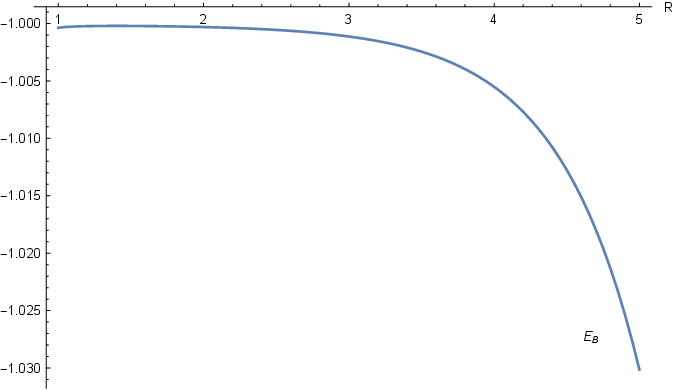}
         \caption{Ground state energy versus $R$.}
         \label{energyvsa1}
     \end{subfigure}
     \hfill
     \begin{subfigure}[b]{0.45\textwidth}
         \centering
         \includegraphics[width=\textwidth]{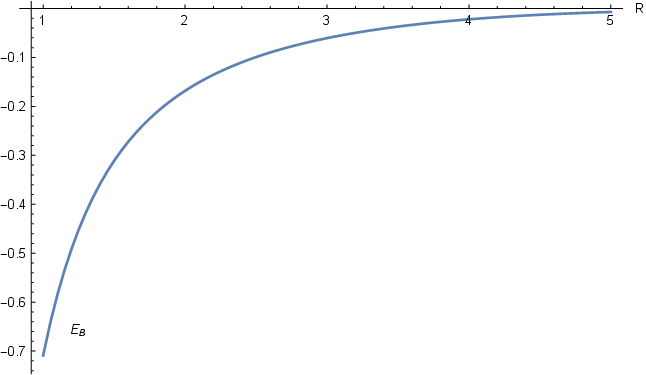}
         \caption{Excited state energy versus $R$.}
         \label{energyvsa2}
     \end{subfigure}
        \caption{Bound state energies $E_B$ versus $R$ for $\lambda=10, a=5.1, \mu=1$ units.}
        \label{fig:ptcircleenergyvsR}
\end{figure}
\end{myremark}

\subsection{Stationary Scattering Problem}
\label{Stationary Scattering Problem for Circle and Point}

Stationary scattering problem for such singular potentials is well-defined, that is, the wave operators $\Omega_{\pm}$ {\it exist and are complete} thanks to Birman-Kuroda theorem \cite{blank2008hilbert}. This theorem states that if the difference between the resolvent  $(H-E)^{-1}$ of  the self-adjoint operator $H$ and the resolvent of the free self-adjoint Hamiltonian $(H_0-E)^{-1}$, this difference being  defined on their common resolvent set,  is  trace class, then wave operators exist and are complete. We have already shown above that $R(E)-R_0(E)$ is trace class. Therefore, the wave operators for defining scattering phenomena exist. Once we have well-defined wave operators, we can study  physically measurable  quantities (e.g., cross section) of a scattering experiment by finding the scattering amplitudes. For this reason, we first need to determine the boundary values of the operator $T(E)$ as $E$ approaches to the positive real axis from above. This is accomplished from the explicit formula of the resolvent written on the complex plane. For convenience, let $E=E_k+ i \epsilon$ where $E_k=k^2$ with $k>0$.  The relation between the resolvent and operator $T$ is given by \cite{Taylor}
\begin{eqnarray}
 R(E)= R_0(E)- R_0(E) T(E) R_0(E) \;. \label{resolventToperator}
\end{eqnarray}
Then, we have the following result for the differential cross section.

\begin{mytheo}
The differential cross section for the rank one perturbation supported by a circle of radius $R$ centered at the origin and by a point at $\mathbf{a}$ outside of the circle is given by
\begin{eqnarray} & & \frac{d \sigma}{d \theta}=|f(\mathbf{k} \rightarrow \mathbf{k}')|^2 = \frac{1}{8 \pi k}\Bigg|
e^{i (\mathbf{k}-\mathbf{k}')\cdot \mathbf{a}} \left(\Phi^{-1}(E_k+i0)\right)_{11} +  J_0(kR) \left(e^{-i \mathbf{k}' \cdot \mathbf{a}} +e^{i \mathbf{k} \cdot \mathbf{a}}  \right) \left(\Phi^{-1}(E_k+i0)\right)_{12} \nonumber \\ & & \hspace{5cm} + \, J_{0}^{2}(kR) \left(\Phi^{-1}(E_k+i0)\right)_{22} \Bigg|^2 \;,
\end{eqnarray}
where $\Phi(E_k+i0)$ is defined by the analytic continuation of (\ref{principalmatrixpointcircle}).
\end{mytheo}

\begin{proof}

Since we have the explicit expression for resolvent (\ref{resolventcirclepoint}) extended onto the complex plane, we can read off the boundary values of  operator $T(E)$ on the positive real axis:
\begin{eqnarray}
 T(E_k+i0)= - \sum_{i,j=1}^{2} |f_i \rangle \left(\Phi^{-1}(E_k+i0)\right)_{ij} \langle f_j| \;, \label{Toperator}
\end{eqnarray}
where
\begin{eqnarray}
 \Phi(E_k+i0)=\left(
\begin{array}{cccc}
\frac{1}{2\pi} \left( - \frac{i \pi}{2} + \log \left(\frac{k}{\mu}\right) \right) & & -\frac{i}{4} H_{0}^{(1)}\left(k a\right) J_0\left(k R\right) \\ \\
-\frac{i}{4} H_{0}^{(1)}\left(k a\right) J_0\left(k R\right) & & \frac1{\lambda_2}- \frac{i}{4} H_{0}^{(1)}\left(k R\right) J_0\left(k R\right) 
\end{array}
\right) \;. \label{Phimatrixonboundary}
\end{eqnarray}
Here we have used $K_0(z)= \frac{i \pi}{2} H_{0}^{(1)}(e^{i \pi/2} z)$ and $I_0(z)=e^{-i \pi/2} J_0(e^{i \pi/2}z)$ for $-\pi < arg(z) < \pi/2$ \cite{Lebedev1965special}. The scattering amplitude denoted by $f$ and the boundary values of the operator $T$ in two dimensions is related by
\begin{eqnarray}
 f(\mathbf{k} \rightarrow \mathbf{k}')= -\frac{1}{4} \sqrt{\frac{2}{\pi k}}  \langle \mathbf{k}' | T(E_k+i0)| \mathbf{k} \rangle \;, \label{scatteringamplitudeTmatrixin2D} 
\end{eqnarray}
where $|\mathbf{k} \rangle$ is the generalized Dirac ket vector and $|\mathbf{k}'|=|\mathbf{k}|$. (We note that there is another choice for the scattering amplitude by ignoring the factor $\sqrt{i/k}$ to get some desirable properties \cite{adhikari1986quantum}, here we use the  conventional version). Substituting the result (\ref{Toperator}) into
\begin{eqnarray}
 \langle \mathbf{k}' | T(E_k+i0)| \mathbf{k} \rangle = \int_{\mathbb{R}^2} \int_{\mathbb{R}^2} e^{i \mathbf{k} \cdot \mathbf{x}-i \mathbf{k}' \cdot \mathbf{x}'} \langle \mathbf{x}'|T(E_k+i0)| \mathbf{x} \rangle \; d^2 x \, d^2 x'
\end{eqnarray}
and using the integral representation of the Bessel function $J_0(x)$ given in (\ref{intrepofbessel1stkind}) we find
\begin{eqnarray} & 
  \langle \mathbf{k}' | T(E_k+i0)| \mathbf{k} \rangle = e^{i (\mathbf{k}-\mathbf{k}')\cdot \mathbf{a}} \left(\Phi^{-1}(E_k+i0)\right)_{11} +  J_0(kR) \left(e^{-i \mathbf{k}' \cdot \mathbf{a}} +e^{i \mathbf{k} \cdot \mathbf{a}}  \right) \left(\Phi^{-1}(E_k+i0)\right)_{12} \nonumber \\ &  + \, J_{0}^{2}(kR) \left(\Phi^{-1}(E_k+i0)\right)_{22}  \;,
\end{eqnarray}
where $\left(\Phi^{-1}(E_k+i0)\right)_{ij}$ is the $ij$th element of the inverse of the matrix $\Phi(E_k+i0)$ given in equation (\ref{Phimatrixonboundary}). 
\end{proof}

\begin{figure}[h!]
    \centering
    \includegraphics{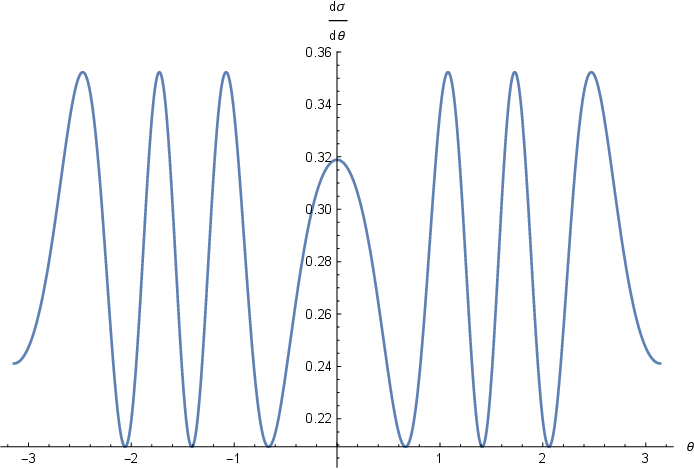}
    \caption{Differential Cross Section versus $\theta$ for $k=2$, $\lambda_2=20$, $a=5$, $R=1$, $\mu=1$ units.}
    \label{fig:scatteringcirclept}
\end{figure}

\begin{figure}[h!]
     \centering
     \begin{subfigure}[b]{0.45\textwidth}
         \centering
         \includegraphics[width=\textwidth]{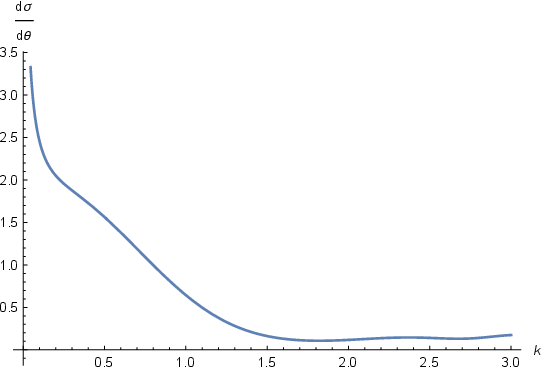}
         \caption{Differential Cross Section versus $k$ for $\theta=0$, $\lambda_2=20$, $a=2$, $R=1$, $\mu=10$ units.}
         \label{diffcrosssectionptcircle1}
     \end{subfigure}
     \hfill
     \begin{subfigure}[b]{0.45\textwidth}
         \centering
         \includegraphics[width=\textwidth]{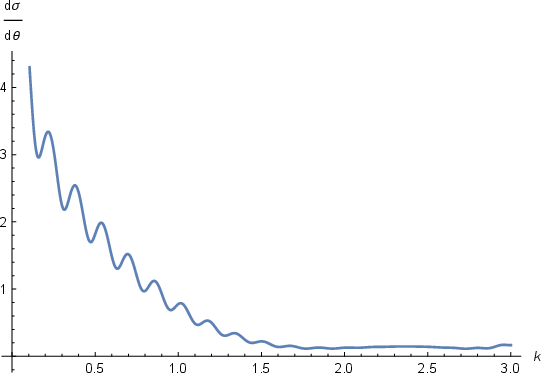}
         \caption{Differential Cross Section versus $k$ for $\theta=0$, $\lambda_2=20$, $a=20$, $R=1$, $\mu=10$ units.}
         \label{diffcrosssectionptcircle2}
     \end{subfigure}
        \caption{Differential Cross Section versus $k$}
        \label{fig:diffcrosssectionptcircle}
\end{figure}

The differential cross section is plotted as a function of $\theta$ in Fig. \ref{fig:scatteringcirclept}. Here we assume that the support of the point defect is at $x=a$ without loss of generality and $\theta$ is the angle between $\mathbf{k}'$ and $\mathbf{k}$ chosen to be along the positive $x$ axis.
One can also plot the differential cross section as a function of $k$ for different choice of parameters, as shown in Figs. \ref{diffcrosssectionptcircle1}  and \ref{diffcrosssectionptcircle2}. The behaviour near $k=0$ of the differential cross section is consistent with the fact that the differential cross section for two-dimensional low energy scatterings blows up with decreasing energy, as emphasized in \cite{landau2013quantum}. 

\section{Rank One Perturbation Supported by a Sphere and a Point in $\mathbb{R}^3$}
\label{Delta Potential Supported by a Sphere and a Point}

In this section, we will consider the rank one perturbation supported by a sphere  and by a point in three dimensions. Since all the techniques and results are similar to the case discussed in the previous section, we will summarize some results without giving  detailed proofs.

The regularized Hamiltonian for this model is given by
\begin{eqnarray} \label{regularizedH2}
    H_{a \Sigma, \,\epsilon}= H_0 - \lambda_1(\epsilon) |\mathbf{a}^{\epsilon}\rangle \langle \mathbf{a}^{\epsilon}| - \lambda_2 |\Sigma^{\epsilon} \rangle \langle \Sigma^{\epsilon} | \;,
\end{eqnarray}
where $\Sigma$ is the sphere centered at the origin with radius $R$ and 
\begin{eqnarray}
    \langle \mathbf{a}^{\epsilon}|\psi \rangle & = &  \int_{\mathbb{R}^3} K_{\epsilon/2}(\mathbf{r}, \mathbf{a}) \psi(\mathbf{r}) \; d^3 r \;, \\ 
     \langle \Sigma^{\epsilon} |\psi \rangle & = & \frac{1}{A(S^2)} \int_{S^2} \left(\int_{\mathbb{R}^3}   K_{\epsilon/2}(\mathbf{r}, \boldsymbol{\sigma}(\theta, \phi)) \psi(\mathbf{r}) \; d^3 r \right) d A \;.
\end{eqnarray}
Here $\boldsymbol{\sigma}:(0,2\pi) \times (0, \pi) \rightarrow S^2$ is the local parametrization given by
\begin{eqnarray}
\boldsymbol{\sigma}(\theta, \phi):=(R \sin \theta \cos \phi, R \sin \theta \sin \phi, R \cos \theta) \;. \label{localchartsphere}
\end{eqnarray}
Proceeding analogously to the previous construction of the resolvent for the  circular defect accompanied by a point defect problem, we obtain the resolvent  essentially in the same form  (\ref{resolventcirclepoint}). In this case, the first diagonal element of the matrix $\Phi$ for $E=-\nu^2$ can be calculated similarly:
\begin{eqnarray}
 \Phi_{11}(-\nu^2) = \lim_{\epsilon \rightarrow 0^+}  \int_{0}^{\infty} K_{t+\epsilon} (\mathbf{a}, \mathbf{a}) \left(e^{-t \mu^2}-e^{-t \nu^2}\right) d t = \frac{(\nu -\mu)}{4 \pi} \;,
\end{eqnarray}
by choosing the bare coupling constant $\lambda_1(\epsilon)$ of the point interaction to be of the same type  as (\ref{barecouplingconstant}) except the heat kernel here is written in three dimensions. Choosing the support of the point defect along the $z$ axis, we find the off-diagaonal matrix elements of $\Phi$ by going to spherical coordinates and evaluating the radial part of the integral by the residue theorem,
\begin{eqnarray}  
 \Phi_{12}(-\nu^2)=\Phi_{21}(-\nu^2)  & = &   -\langle \mathbf{a}|R_0(-\nu^2)|\Sigma \rangle = - \int_{\mathbb{R}^3} \frac{e^{i \mathbf{p} \cdot \mathbf{a}}}{(p^2 + \nu^2)}\;\frac{\sin (p R)}{p R} \frac{d^3 p}{(2\pi)^3}  \nonumber \\ & = & - \frac{1}{4\pi \nu a R} \; e^{-\nu a} \sinh (\nu R) \label{Phioffdiagonal1spherept}\;,
\end{eqnarray}
where we have used
\begin{eqnarray}
\langle \mathbf{p} |\Sigma \rangle =\frac{\sin (p R)}{p R} \;.
\end{eqnarray}
Similarly, 
\begin{eqnarray}
 \Phi_{22}(-\nu^2) &=&  \frac{1}{\lambda_2} - \langle \Sigma | R_0(-\nu^2)| \Sigma \rangle = \frac{1}{\lambda_2}- \int_{\mathbb{R}^3} \frac{1}{p^2+\nu^2} \frac{\sin^2(p R)}{(p R)^2} \frac{d^3 p}{(2\pi)^3}\nonumber \;, \\
 & = &  \frac{1}{\lambda_2} - \frac{1}{4 \pi \nu R^2} e^{-\nu R} \sinh(\nu R) \;. \label{Philastdiagonalsphere1}
\end{eqnarray}
The resolvent of the model is formally given by equation (\ref{resolventcirclepoint}), where $|f_2 \rangle =|\Sigma\rangle$ and the matrix $\Phi$ can be defined on the complex plane by an analytic continuation of the above expressions. It is easy to see that the matrix elements of  above matrix looks similar to our two-dimensional version if we express  its entries  in terms of the Bessel functions using $I_{1/2}(z)=\sqrt{\frac{2}{\pi z}}\sinh z$ and $K_{1/2}(z)=\sqrt{\frac{\pi}{2z}}e^{-z}$,
\begin{eqnarray}
\Phi(-\nu^2) =\left(
\begin{array}{cccc}
 \frac{1}{4\pi}(\nu-\mu) &  & - \frac{1}{4\pi \sqrt{a R}} \; K_{1/2}(\nu a) \; I_{1/2}(\nu R)  \\ \\
- \frac{1}{4\pi \sqrt{a R}} \; K_{1/2}(\nu a) \; I_{1/2}(\nu R) & & \frac{1}{\lambda_2} - \frac{1}{4\pi R} K_{1/2}(\nu R) \; I_{1/2}(\nu R)
\end{array}
\right) \;. \label{Phispherepointnew}
\end{eqnarray}

\subsection{Bound State Problem}
\label{Bound State Problem for Sphere and Point}

Bound state analysis of this problem is performed exactly in the same manner as in the case of rank one perturbation supported by a circle and a point. For this reason, we are not going to derive the analogous expressions for the flow of the eigenvalues with respect to $\nu$. Positivity of the flow of eigenvalues equally holds in this case so we conclude that there are at most two bound states (and at least one bound state).

\begin{figure}[h!]
     \centering
     \begin{subfigure}[b]{0.45\textwidth}
         \centering
         \includegraphics[width=\textwidth]{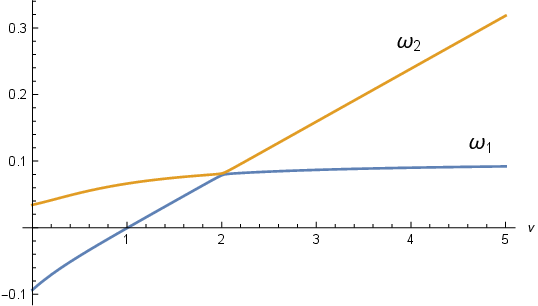}
         \caption{Eigenvalues of the principal matrix $\Phi$ versus $\nu$ for $\lambda_2=10$, $a=2$, $R=1$, and $\mu=1$ units.}
    \label{fig:eigenvaluesspherept1}
     \end{subfigure}
     \hfill
     \begin{subfigure}[b]{0.45\textwidth}
         \centering
         \includegraphics[width=\textwidth]{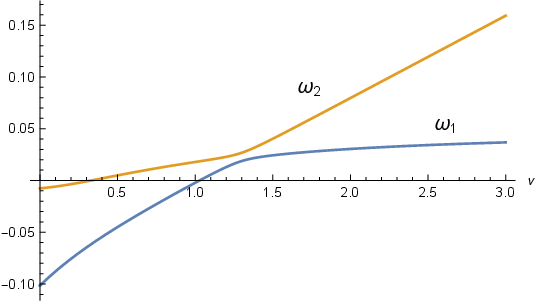}
        \caption{Eigenvalues of the principal matrix $\Phi$ versus $\nu$ for $\lambda_2=20$, $a=2$, $R=1$, and $\mu=1$ units.}
    \label{fig:eigenvaluesspherept2}
     \end{subfigure}
        \caption{Eigenvalues of $\Phi$ versus $\nu$}
        \label{fig:eigenvaluessphere}
\end{figure}

One can plot the eigenvalues as a function of $\nu$ for particular values of the parameters. As shown in the previous section, zeroes $\nu_*$ of the eigenvalues correspond to the bound state energies $E=-\nu_{*}^2$. It is interesting to notice that there is only one bound state if we choose the same values of the parameters for the circular defect perturbed by a point defect problem, as shown in Fig. \ref{fig:eigenvaluesspherept1}. The reason for this may be based on the fact that particle has more freedom to escape from the spherical defect compared to the circular defect.   
If we increase the strength of the spherical defect, the second bound state appears as shown in Fig. \ref{fig:eigenvaluesspherept2}.

\begin{figure}[h!]
     \centering
     \begin{subfigure}[b]{0.45\textwidth}
         \centering
         \includegraphics[width=\textwidth]{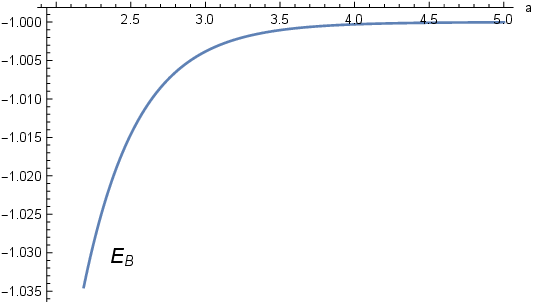}
         \caption{Ground state energy versus $a$}
         \label{fig:sphereenergyvsa1}
     \end{subfigure}
     \hfill
     \begin{subfigure}[b]{0.45\textwidth}
         \centering
         \includegraphics[width=\textwidth]{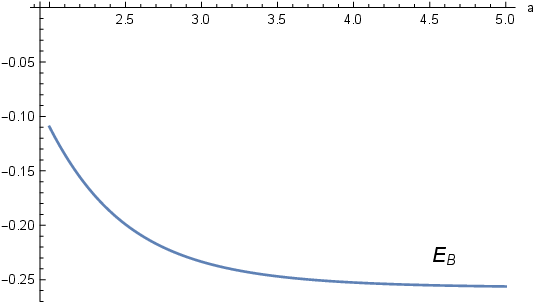}
         \caption{Excited state energy versus $a$}
         \label{fig:sphereenergyvsa2}
     \end{subfigure}
        \caption{Bound state energies $E_B$ versus $a$ for $\lambda_2=20, R=1, \mu=1$ units.}
        \label{fig:ptsphereenergyvsa}
\end{figure}

\begin{figure}[h!]
     \centering
     \begin{subfigure}[b]{0.45\textwidth}
         \centering
         \includegraphics[width=\textwidth]{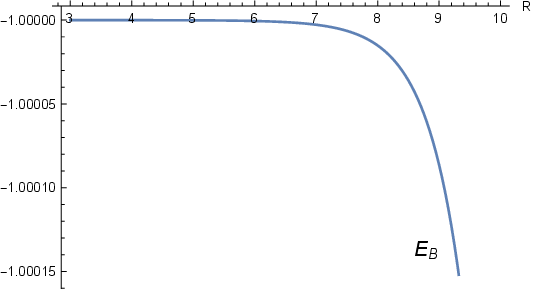}
         \caption{Ground state energy versus $R$}
         \label{fig:sphereenergyvsR1}
     \end{subfigure}
     \hfill
     \begin{subfigure}[b]{0.45\textwidth}
         \centering
         \includegraphics[width=\textwidth]{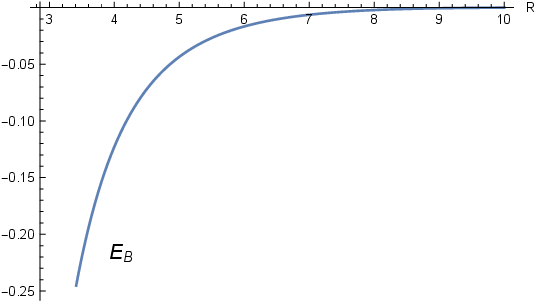}
         \caption{Excited state energy versus $R$}
         \label{fig:sphereenergyvsR2}
     \end{subfigure}
        \caption{Bound state energies versus $R$ for $\lambda_2=150, a=10.1, \mu=1$ units.}
        \label{fig:ptsphereenergyvsR}
\end{figure}

One can find how the bound state energies change with respect to the parameters $R$ and $a$ by numerically solving the zeroes of the eigenvalues $\omega_1$ and $\omega_2$. They are plotted in Figs. \ref{fig:ptsphereenergyvsa}  and \ref{fig:ptsphereenergyvsR}.
By following the same line of arguments, we have the following theorem.
\begin{mytheo} Let $\mathbf{a} \in \mathbb{R}^3$ and $\Sigma$ be the sphere centered at the origin with radius $R<a$. Then, the essential spectrum of $H_{a \Sigma}$ associated with rank one perturbation supported by $\Sigma$ and a point coincides with the essential spectrum of the free Hamiltonian, i.e., $\sigma_{ess}(H_{a \Sigma})=\sigma_{ess}(H_0)=[0, \infty)$. Furthermore, the point spectrum $\sigma_p(H_{a \Sigma})$ of $H_{a \Sigma}$ lies in the negative real axis and $H_{a \Sigma}$ has at most two negative eigenvalues (counting multiplicity) and always has  one. Let $\Real(k)=0$ and $\Imaginary(k)>0$, then $k^2 \in \sigma_p(H_{a \Sigma})$ if and only if $\det \Phi(k^2)=0$ and multiplicity (degeneracy) of the eigenvalue $k^2$ is the same as the multiplicity of the zero eigenvalue of the matrix $\Phi(k^2)$. Moreover, let $E=-\nu_{*}^2<0$ be an eigenvalue of $H_{a \Sigma}$, then the eigenfunction $|\psi_{ev}\rangle$ associated with this eigenvalue is given by
\begin{eqnarray*}
\psi_{ev}(\mathbf{r})= \sum_{i=1}^{2} \langle \mathbf{r}| R_0(-\nu_{*}^2)|f_i \rangle A_i \;,
\end{eqnarray*}
where $(A_1, A_2)$ is an eigenvector with zero eigenvalue of $\Phi(-\nu_{*}^2)$ and $|f_1 \rangle =|\mathbf{a} \rangle$, $|f_2 \rangle = |\Sigma \rangle$.
\end{mytheo}

\subsection{Stationary Scattering Problem}
\label{Stationary Scattering Problem for Sphere and Point}

\begin{figure}[h!]
    \centering
    \includegraphics{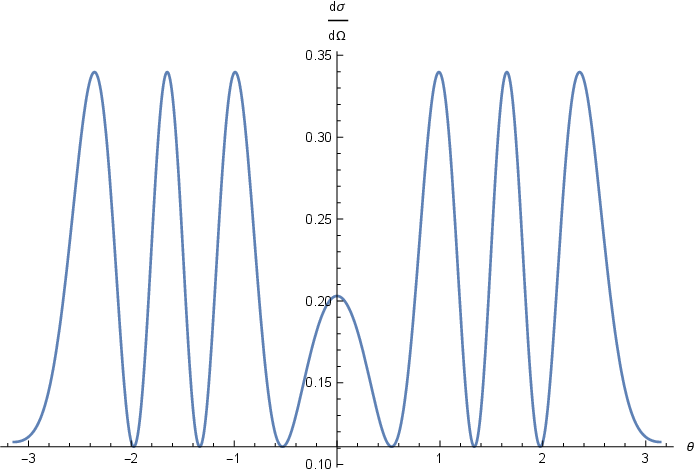}
    \caption{Differential cross section versus $\theta$ for $k=2$, $\lambda_2=10$, $a=5$, $R=1$, $\mu=1$ units.}
    \label{fig:diffcrosssectionspherept}
\end{figure}

For the scattering problem, we similarly find the boundary values of the principal operator by analytical continuation   
\begin{eqnarray}
 \Phi(E_k+i0)=\left(
\begin{array}{cccc}
\frac{1}{4\pi} \left( -i k - \mu \right) & & -\frac{1}{4 \pi a R k} e^{i k a} \; \sin (k R) \\ \\
-\frac{1}{4 \pi a R k} e^{i k a} \; \sin (k R)  & & \frac1{\lambda_2}- \frac{e^{i k R}}{4 \pi R^2 k} \sin (k R) \\
\end{array}
\right) \;, \label{Phimatrixonboundaryspherept}
\end{eqnarray}
and
\begin{eqnarray} & & \hskip-1cm
\langle \mathbf{k}' | T(E_k+i0)| \mathbf{k} \rangle =  - \sum_{i,j=1}^{2} \langle \mathbf{k}'|f_i \rangle \left(\Phi^{-1}(E_k+i0)\right)_{ij} \langle f_j| \mathbf{k} \rangle \nonumber \\ & &  =  - \bigg(e^{i (\mathbf{k}-\mathbf{k}')\cdot \mathbf{a}}  \left(\Phi^{-1}(E_k+i0)\right)_{11} + \frac{\left(e^{-i \mathbf{k}' \cdot \mathbf{a}} + e^{i \mathbf{k} \cdot \mathbf{a}}\right) \sin (k R)}{k R} \; \left(\Phi^{-1}(E_k+i0)\right)_{12} \nonumber \\ & & \hspace{2cm} + \,   \frac{\sin^2 (k R)}{k^2 R^2} \; \left(\Phi^{-1}(E_k+i0)\right)_{22}  \bigg) \;.
\end{eqnarray}
Hence, we find the scattering amplitude from the formula $f(\mathbf{k} \rightarrow \mathbf{k}')=-\frac{1}{4 \pi}\langle \mathbf{k}' | T(E_k+i0)| \mathbf{k} \rangle  $, and the graph of the differential cross section $\frac{d \sigma}{d \Omega}=|f(\mathbf{k} \rightarrow \mathbf{k}')|^2 $ as a function of $\theta$ is given in Fig. \ref{fig:diffcrosssectionspherept}.
Let us summarize the result.
\begin{mytheo}
The differential cross section for the rank one perturbation supported by a sphere of radius $R$ centered at the origin and by a point at $\mathbf{a}$ outside of the sphere is given by
\begin{eqnarray} & & \frac{d \sigma}{d \Omega}=|f(\mathbf{k} \rightarrow \mathbf{k}')|^2 = \frac{1}{16 \pi^2} \bigg|e^{i (\mathbf{k}-\mathbf{k}')\cdot \mathbf{a}}  \left(\Phi^{-1}(E_k+i0)\right)_{11} + \frac{\left(e^{-i \mathbf{k}' \cdot \mathbf{a}} + e^{i \mathbf{k} \cdot \mathbf{a}}\right) \sin (k R)}{k R} \; \left(\Phi^{-1}(E_k+i0)\right)_{12} \nonumber \\ & & \hspace{4cm} + \,   \frac{\sin^2 (k R)}{k^2 R^2} \; \left(\Phi^{-1}(E_k+i0)\right)_{22}  \bigg|^2 \;,
\end{eqnarray}
where $\Phi(E_k+i0)$ is given by (\ref{Phimatrixonboundaryspherept}).
\end{mytheo}

\begin{figure}[h!]
     \centering
     \begin{subfigure}[b]{0.45\textwidth}
         \centering
         \includegraphics[width=\textwidth]{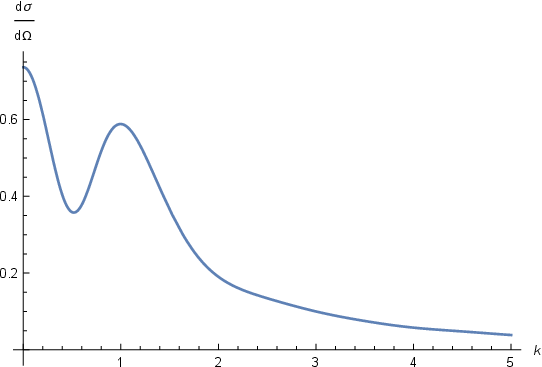}
        \caption{Differential cross section versus $k$ for $\theta=0$, $\lambda_2=5$, $a=2$, $R=1$, $\mu=1$ units.}
    \label{fig:diffcrosssectionsphereptvsk1}
     \end{subfigure}
     \hfill
     \begin{subfigure}[b]{0.45\textwidth}
         \centering
         \includegraphics[width=\textwidth]{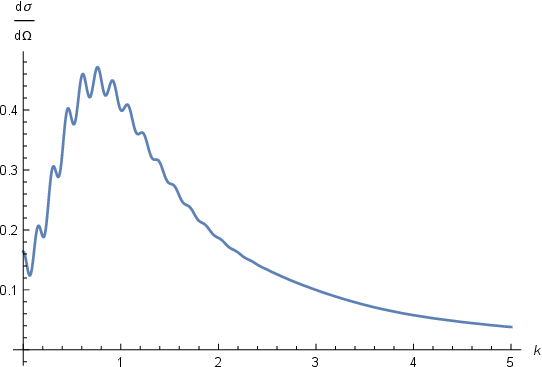}
        \caption{Differential cross section versus $k$ for $\theta=0$, $\lambda_2=20$, $a=2$, $R=1$, $\mu=1$ units.}
    \label{fig:diffcrosssectionsphereptvsk2}
     \end{subfigure}
        \caption{Differential cross section versus $k$.}
        \label{fig:diffcrosssectionsphereptvsk}
\end{figure}

For the forward scattering, the differential cross section is plotted as a function of $k$ for the below values of the parameters, as shown in Figs. \ref{fig:diffcrosssectionsphereptvsk1}  and \ref{fig:diffcrosssectionsphereptvsk2}.

\section{Small Deformations of a Circle in $\mathbb{R}^2$}
\label{Small Deformations of a Circle}

It would be interesting to ask how the bound state spectrum and scattering cross section for the above (or similar) problems change under small deformation of the support of the rank one perturbations. Let us first briefly define the normal deformations of a general curve in two dimensions. We consider a regular planar  curve $\Gamma$ parametrized with its arc length $s$ with finite length. The Serret-Frenet equations for this curve are given by
$\mathbf{t}=\frac{d \boldsymbol{\gamma}}{ds}$, $\frac{d \mathbf{t}}{d s}=\kappa \mathbf{n}$, and $\frac{d \mathbf{n}}{d s}= - \kappa \mathbf{t}$, where $\mathbf{t}$, $\mathbf{n}$ are the tangent and normal vectors and $\kappa$ is the curvature of the curve $\Gamma$ \cite{do2016differential}.
The small deformation along a normal direction of the curve $\Gamma$ is defined by
\begin{equation}
\tilde{\boldsymbol{\gamma}}(s)= \boldsymbol{\gamma}(s)+ \epsilon h(s) \mathbf{n} (s) \;, \label{deformationofcurve}
\end{equation}
where $h$ is assumed to be a smooth function of $s$. It is worth pointing out that $\epsilon$ here is a small deformation parameter, \textit{not the same parameter used for regularization}.

The length of the deformed curve $\tilde{\Gamma}$ up to order $\epsilon$ is given by
\begin{eqnarray}
 L(\tilde{\Gamma}) & = &  \int_{0}^{L} \frac{d \tilde{s}}{ds} \; d s = \int_{0}^{L} \left( \frac{d \boldsymbol{\tilde{\gamma}}}{ds} \cdot \frac{d \boldsymbol{\tilde{\gamma}}}{ds} \right)^{1/2}d s = \int_{0}^{L} \left(\left(1-\epsilon \kappa(s)h(s)\right)^2 + \epsilon^2 \left(\frac{d h(s)}{ds}\right)^2\right)^{1/2} \; d s
 \nonumber \\ & = & \int_{0}^{L} \left( 1- \epsilon \kappa(s) h(s)+ O(\epsilon^2) \right) \; d s = L(\Gamma)- \epsilon \int_{0}^{L} \kappa(s)h(s) ds +O(\epsilon^2) \;.
\end{eqnarray}
If $\Gamma$ is a circle of radius $R$, $\kappa=1/R$ so that 
\begin{eqnarray}
L(\tilde{\Gamma})=2\pi R - \frac{\epsilon}{R} \int_{0}^{L} h(s) ds + O(\epsilon^2) \;. \label{lengthofdeformedcircle}
\end{eqnarray} 
When we do reparametrization of the curve by the angle $\theta$, we will use the same notation for the functions $h$, $\boldsymbol{\gamma}$, and $\mathbf{n}$.

\subsection{Perturbative First Order Calculation of the Bound State Energy}
\label{First Order Calculation of the Bound State Energy for Deformed Circle}

We consider here that the interaction is formally represented by $\lambda |\tilde{\Gamma} \rangle \langle \tilde{\Gamma}|$ - rank one perturbation   supported on a deformed circle.
Since the support of the defect has co-dimension one, the renormalization is not required for this model and the resolvent of the Hamiltonian $H$ associated with a deformed circular defect can be found by using similar arguments summarized previously, as a result  we find 
\begin{eqnarray}
 R(E)= R_0(E)+R_0(E) | \tilde{\Gamma} \rangle \frac{1}{\tilde{\Phi}(E)} \langle \tilde{\Gamma}| R_0(E) \;, \label{resolventofdeformedcircle}
\end{eqnarray}
where we denote the deformation of the circle $\tilde{S^{1}}$ by $\tilde{\Gamma}$ for notational simplicity. For the bound state, we need to calculate
\begin{eqnarray}
 \tilde{\Phi}(-\nu^2)= \frac{1}{\lambda} - \langle \tilde{\Gamma} | R_0(-\nu^2) | \tilde{\Gamma} \rangle = \frac{1}{\lambda} -  
 \int_{\mathbb{R}^2} \frac{|\langle \tilde{\Gamma}|\mathbf{p} \rangle|^2}{p^2+\nu^2}\; \frac{d^2 p}{(2\pi)^2} \;. \label{Phideformedcircle1}
\end{eqnarray}
Using
\begin{eqnarray}
 \langle \tilde{\Gamma}|\mathbf{p} \rangle = \frac{1}{L(\tilde{\Gamma})} \int_{0}^{L} e^{i \mathbf{p}\cdot \boldsymbol{\tilde{\gamma}}(s)} \; |\boldsymbol{\tilde{\gamma}'}(s)| \, d s \;,
\end{eqnarray}
and expanding the exponential $e^{i \epsilon h(\theta) \mathbf{p} \cdot \mathbf{n}(\theta)}$ in $\epsilon$ and the fact $|\boldsymbol{\tilde{\gamma}'}(s)|= 1-\frac{\epsilon}{R}h(s) + O(\epsilon^2)$, it is easy to show that
\begin{eqnarray} & & \hskip-2cm
\Tilde{\Phi}(-\nu^2) =  \frac{1}{\lambda} - \frac{1}{(2\pi)^2} \left( 1 + \frac{\epsilon}{\pi R} \int_{0}^{2\pi} h(\theta) d \theta \right)
\Bigg[ \int_{\mathbb{R}^2} \bigg( \int_{0}^{2\pi} \int_{0}^{2\pi} \frac{e^{i \mathbf{p} \cdot (\boldsymbol{\gamma}(\theta_1)-\boldsymbol{\gamma}(\theta_2)) }}{p^2 + \nu^2} \; \nonumber \\ & & \hskip-1cm \times \bigg(1- \frac{\epsilon}{R} (h(\theta_1)+h(\theta_2)) + i \epsilon ((\mathbf{p} \cdot \mathbf{n}(\theta_1))h(\theta_1) - (\mathbf{p} \cdot \mathbf{n}(\theta_2))h(\theta_2) )  \bigg) d \theta_1 d \theta_2 \bigg) \Bigg] \frac{d^2 p}{(2\pi)^2} + O(\epsilon^2)  \;. \label{defomedcirclephiexpanded}
\end{eqnarray}
Let us consider the first integral in the square bracket above:
\begin{eqnarray}
\int_{\mathbb{R}^2} \left( \int_{0}^{2\pi} \int_{0}^{2\pi} \frac{e^{i \mathbf{p} \cdot (\boldsymbol{\gamma}(\theta_1)-\boldsymbol{\gamma}(\theta_2)) }}{p^2 + \nu^2} d \theta_1 d \theta_2 \right) \frac{d^2 p}{(2\pi)^2} \;.
\end{eqnarray}
The uniformly convergent plane wave expansion in two dimensions \cite{adhikari1986quantum}
\begin{eqnarray}
e^{i \mathbf{p} \cdot \mathbf{r}} = \sum_{m=0}^{\infty} \varepsilon_m i^m J_m(p r) \cos(m \theta) \;, \label{planewaveexpansion2d}
\end{eqnarray}
with $\varepsilon_0=1$, $\varepsilon_m=2$ if $m>0$, and $\theta$ being the angle between $\mathbf{p}$ and $\mathbf{r}$ helps us to compute the above integral with respect to the angle variables easily and left with the integration over the variable $p$ only:  
\begin{eqnarray}
(2\pi) \int_{0}^{\infty} \frac{J_{0}^{2}(p R)}{p^2 + \nu^2} \; p \, d p \;, \end{eqnarray}
where we have used $\int_{0}^{2\pi} \cos(m (\theta-\theta_k)) d \theta=2\pi \delta_{m0}$. Thanks to the integral representation \cite{gradshteyn2014table} 
\begin{eqnarray}
\int_{0}^{\infty} \frac{x}{x^2+a^2} J_{0}^{2}(x) d x = I_0(a)K_0(a) \;, \label{I0K0integral}
\end{eqnarray} 
we find
\begin{eqnarray}
\int_{\mathbb{R}^2} \left( \int_{0}^{2\pi} \int_{0}^{2\pi} \frac{e^{i \mathbf{p} \cdot (\boldsymbol{\gamma}(\theta_1)-\boldsymbol{\gamma}(\theta_2)) }}{p^2 + \nu^2} d \theta_1 d \theta_2 \right) \frac{d^2 p}{(2\pi)^2} = (2 \pi) I_{0}(\nu R) K_{0}(\nu R) \;. \label{1stintegralPhideformedcircle} 
\end{eqnarray}
For the second integral in equation (\ref{defomedcirclephiexpanded}), it is sufficient to consider
\begin{eqnarray}
\int_{\mathbb{R}^2} \left( \int_{0}^{2\pi} \int_{0}^{2\pi} \frac{e^{i \mathbf{p} \cdot (\boldsymbol{\gamma}(\theta_1)-\boldsymbol{\gamma}(\theta_2)) }}{p^2 + \nu^2} h(\theta_1) d \theta_1 d \theta_2 \right) \frac{d^2 p}{(2\pi)^2}  \;.
\end{eqnarray}
With the help of the plane wave expansion (\ref{planewaveexpansion2d}) and the formula (\ref{I0K0integral}), the above integral becomes
\begin{eqnarray}
I_0(\nu R) K_0(\nu R) \left( 
\int_{S^1} h(\theta) d \theta \right) \;. \label{deformedPhicircle2}\end{eqnarray}
The last integral in (\ref{defomedcirclephiexpanded}) can be computed similarly by first rewriting the expression $i (\mathbf{p} \cdot \mathbf{n}(\theta)) e^{i \mathbf{p} \cdot \boldsymbol{\gamma}(\theta)} = \frac{\partial}{\partial R} (e^{i \mathbf{p} \cdot \boldsymbol{\gamma}(\theta)})$ and $\frac{d J_0(x)}{d x}=-J_1(x)$ we find
\begin{eqnarray} & & \hskip-2cm
\int_{\mathbb{R}^2} \bigg( \int_{0}^{2\pi} \int_{0}^{2\pi} \frac{e^{i \mathbf{p} \cdot (\boldsymbol{\gamma}(\theta_1)-\boldsymbol{\gamma}(\theta_2)) }}{p^2 + \nu^2} \; i \, (\mathbf{p} \cdot \mathbf{n}(\theta_1))h(\theta_1) d \theta_1 d \theta_2 \bigg)  \frac{d^2 p}{(2\pi)^2} \nonumber \\ & & = 
- \left( \int_{S^1} h(\theta) d \theta \right) \int_{0}^{\infty} J_0(p R) J_1(p R) \, \frac{p^2}{p^2+\nu^2} \; d p \;.
\end{eqnarray}
Rewriting $\frac{p^2}{p^2 +\nu^2}$ as $1-\frac{\nu^2}{p^2+\nu^2}$, and using the formula (6.512) in \cite{gradshteyn2014table},
\begin{eqnarray}
\int_{0}^{\infty} J_{\nu}(\alpha x) J_{\nu-1}(\alpha x) d x = \frac{1}{2 \alpha} \;, \label{intofJnuJnu-1}
\end{eqnarray}
and formula (6.577) in \cite{gradshteyn2014table}, 
\begin{eqnarray}
\int_{0}^{\infty} \frac{J_0 (p R) J_1(p R)}{p^2 + \nu^2} d p = \frac{1}{\nu} I_{1}(\nu R) K_0(\nu R) \;,
\end{eqnarray}
it follows that 
\begin{eqnarray} & & \hskip-3cm
\int_{\mathbb{R}^2} \bigg( \int_{0}^{2\pi} \int_{0}^{2\pi} \frac{e^{i \mathbf{p} \cdot (\boldsymbol{\gamma}(\theta_1)-\boldsymbol{\gamma}(\theta_2)) }}{p^2 + \nu^2} \; i \, (\mathbf{p} \cdot \mathbf{n}(\theta_1))h(\theta_1) d \theta_1 d \theta_2 \bigg)  \frac{d^2 p}{(2\pi)^2} \nonumber \\ & & =-\left( \frac{1}{2 R} - \nu I_1 (\nu R) K_0(\nu R) \right) \left( \int_{0}^{2\pi} h(\theta) d \theta \right) \;. \label{deformedPhicircle3}
\end{eqnarray}
After combining all the above results (\ref{1stintegralPhideformedcircle}), (\ref{deformedPhicircle2}), and (\ref{deformedPhicircle3}), we finally obtain
\begin{eqnarray} & & \hskip-1.5cm
\Tilde{\Phi}(-\nu^2) = \frac{1}{\lambda} - \frac{1}{2\pi} I_0(\nu R) K_0(\nu R) + \frac{\epsilon}{2 \pi^2} \left( - \frac{1}{2R} + \nu I_0(\nu R) K_1(\nu R) \right)  \left(\int_{0}^{2\pi} h(\theta) d \theta \right) + O(\epsilon^2) \;,\label{deformedcirclePhiforboundstate}
\end{eqnarray}
where we have used $I_1(x)K_0(x)+I_0(x)K_1(x)=1/x$.

When there is no deformation ($\epsilon=0$), we have only one bound state. This can be seen by simply expressing the second term $I_0(\nu R) K_0(\nu R)$ using its integral representation (\ref{I0K0integral}):
\begin{eqnarray}
\frac{1}{\lambda} = \frac{1}{2\pi}  I_0(\nu R) K_0(\nu R) = \frac{1}{2\pi} \int_{0}^{\infty} \frac{x}{x^2+\nu^2 R^2} J_{0}^{2}(x) d x \;.
\end{eqnarray}
Then, by taking the derivative of the right hand side with respect to $\nu$ under the integral sign, it is easy to see that the right hand side of the above equation is a decreasing function of $\nu$ for given parameters $\lambda$ and $R$. Therefore, there is a unique solution, say  $\nu_0$, to the above equation.

It is important to notice that deformations satisfying $\int_{0}^{2\pi} h(\theta) d \theta =0$ do not change the bound state energies up to first order in $\epsilon$. Since we evaluate the deformation to order $\epsilon$ we can actually solve the bound state energy for the deformed curve to the same order. In \cite{pointinteractionsonmanifolds2, erman2019perturbative} we derived a general formula for perturbations of eigenvalues for small perturbations of the principal matrix $\Phi$, here we have a one-dimensional version of this formula so we can directly use the expansion above. Let $\nu= \nu_0 + \epsilon \nu_1 + O(\epsilon^2)$, where $\nu_0$ denotes the solution to the original unperturbed circle case. Then, the bound state energy $E_B=-(\nu_0+\epsilon \nu_1)^2$ for the deformed circular defect can be found by the zeroes of $\tilde{\Phi}$.  This is achieved up to order $\epsilon$ by simply expanding its first term around $\nu_0$, 
\begin{equation}
\frac{1}{\lambda} - \frac{1}{2\pi}  I_0((\nu_0+\epsilon \nu_1) R) K_0((\nu_0+ \epsilon \nu_1) R) +  \frac{\epsilon}{2 \pi^2} \left( - \frac{1}{2R} + \nu_0 I_0(\nu_0 R) K_1(\nu_0 R) \right)  \left(\int_{0}^{2\pi} h(\theta) d \theta \right) =0 \;, \end{equation}
and using the fact that the zeroth order term cancels out $\frac{1}{\lambda}$ above to get the solution $\nu_1$. Hence, we obtain an explicit formula for the bound state energy up to order $\epsilon$
\begin{eqnarray}
E_B = -\nu_{0}^{2} - \epsilon \, \frac{2 \nu_0}{\pi R} \; \left( \frac{\left( \frac{1}{2 R} - \nu_0 I_0(\nu_0 R) K_1(\nu_0 R)\right)}{I_1(\nu_0 R) K_0(\nu_0 R) - I_0(\nu_0 R) K_1(\nu_0 R)}\right) \left( \int_{0}^{2\pi} h(\theta) d \theta \right) + O(\epsilon^2) \;  
,\end{eqnarray}
which can be further simplified into the following.
\begin{mytheo} Under the small deformation of the circle described by (\ref{deformationofcurve}), the bound state energy of the system up to first order in $\epsilon$ is given by  
\begin{eqnarray}
E_B = -\nu_{0}^{2} - \epsilon \, \frac{ \nu^2_0}{\pi R} \left( \int_{0}^{2\pi} h(\theta) d \theta \right) + O(\epsilon^2) \;.  
\end{eqnarray}
\end{mytheo}
The simplicity of the first order result is remarkable, and hints at a geometric interpretation.
Suppose that instead of the original circle with radius $R$ we replace the circle with a circle of radius $R-\epsilon \langle h \rangle$ where 
$\langle h \rangle =\frac{1}{2 \pi R}\int_{0}^{2\pi} h(\theta) R d\theta$ (note that the normal in the curvature description is inward). Because we are now using  a delta function supported by a circle, we do have the same eigenvalue equation,
\begin{equation}
\frac{1}{\lambda} - \frac{1}{2\pi}  I_0 \big((\nu_0+\epsilon \nu_1) (R-\epsilon \langle h \rangle)\big) K_0 \big((\nu_0+ \epsilon \nu_1) (R-\epsilon \langle h \rangle)\big)=0 \;.
\end{equation} 
If we expand all the terms to order $\epsilon$, we find the relation
$R\nu_1=\nu_0 \langle h \rangle$.
By using $E_B=-(\nu_0+\epsilon \nu_1)^2=-\nu_0^2-2\epsilon \nu_0 \nu_1$ we find exactly the above result. 
So we state this observation as follows.
\begin{mycorollary}
A small deformation in the normal direction  of a given circle, which supports an attractive delta function, leads to a perturbation of the original  bound state energy,  to   first order the resulting change can be obtained as follows:  increase the initial radius  by  an amount equal to the average of the deformation over the given  circle,  then compute  the first order perturbation of the  bound state  energy corresponding to  this new circle with the same coupling constant. 
\end{mycorollary}

\begin{myremark}
It is tempting to push this to the second order and  search for, if there is any, a geometric interpretation of the result. But the calculations are rather involved so we postpone it for future work. Note that the circle problem per se  can be solved by elementary methods, that is by choosing polar coordinates at the center and writing the delta function along the radial direction. However, a general curve cannot be solved by this approach as there is no natural coordinate system to choose. In the case of small deformation, one can think of rank one perturbation supported on this curve as a rank one perturbation supported on the original circle plus a series of perturbations. This idea leads to, even to first order, a term of the form of $\epsilon \frac{d \delta (r-R)}{d r}\int_{0}^{2\pi} h(\theta) d\theta $ and  some additional ones coming from the change of arc-length  as well as the change of total length.
Here  the derivative of delta function term is important since the wave function is of the form (disregarding the normalization)
$$
I_0( r\nu_0)K_0(R\nu_0) \theta(R-r) +I_0(R\nu_0)K_0(r\nu_0)\theta(r-R),
$$
and the usual first order perturbation of energy, which is found by evaluating the expectation value in the state of interest, leads to a 
divergence (here we need to use the symmetric choice for the theta function as often used in  distribution theory).
\end{myremark}

\begin{figure}[h!]
    \centering
    \includegraphics{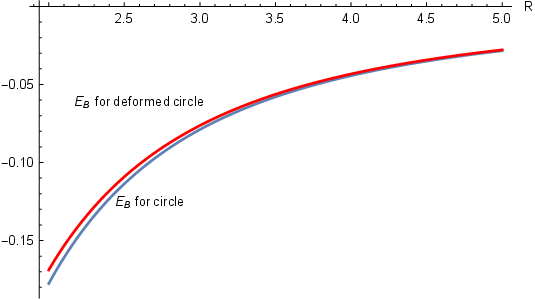}
    \caption{Bound state energy for the circular defect and for the deformed circular defect versus $R$ with $\epsilon=0.1$, $\lambda=10$ units.}
    \label{fig:boundstatedeformedcircle}
\end{figure}

The single bound state energy $E_B$ for the original circular defect can numerically be plotted as a function of $R$ with fixed values of $\lambda$. For a particular deformation $h(\theta)=h_0 \sin^2 \theta$ with $h_0=1$ unit, we can plot how the bound state energy $E_{B}$ changes with respect to $R$ numerically with the help of Mathematica, as shown in Fig. \ref{fig:boundstatedeformedcircle}.    
For a given $R$, it is easy to see that the function $\tilde{\Phi}$ is a decreasing function of $\lambda$ for all $\nu>0$. This implies that the bound state energies decrease with increasing strength $\lambda$, as expected.

\begin{myremark}
In \cite{exner2001geometrically}, the authors considered an infinite curve in the plane as the support of the delta potential and give a precise meaning to the formal Hamiltonian via introducing a locally orthogonal system in the vicinity of the curve $\Gamma$ under some assumptions and curvilinear coordinates are given by $\mathbf{x}(s,u)= \boldsymbol{\gamma}(s)+ u \mathbf{n}(s)$, which is similar to our deformation formula (\ref{deformationofcurve}). However, this expression has a  rather different purpose: Given any function in $L^{\infty}(-1,1)$, one defines a family of  scaled potentials in the straightened strip conveniently expressed in the  local coordinates  given above. Then, resolvents of the Hamiltonians associated with these scaled potentials are constructed. Finally, one shows that Hamiltonians associated with the scaled potentials converge to the formal Hamiltonian associated with a delta potential supported on $\Gamma$ in the norm resolvent sense. Our aim here is to study how the deformation of the rank one potential supported on the circle changes the spectrum of the problem.
\end{myremark}
\begin{myremark}
A slightly different version of our deformed circle problem, in which the discrete spectrum of delta potential supported by a circle with a varying coupling constant, in particular a step function on the circle, has been studied in \cite{exner2003spectra}. The problem of the ring delta potential with variable coupling constant can be considered as an equivalent problem of delta function supported by a deformed circle at first glance. Notice, for instance, that given $h(\theta)$ one may find a function $\alpha(\theta)$ such that $\delta(r-(R +\epsilon h(\theta)))= \alpha(\theta) \delta(r-R)$. However, under the transformation $r'=r-\epsilon h(\theta)$ and $\theta'=\theta$, the  Laplacian must also be transformed, which makes the problem complicated and not quite identical to the usual free Hamiltonian. \end{myremark}

\begin{myremark} There are also works on more general curves chosen for  the support of the delta functions. For instance, the asymptotic behaviour of the bound state energies of the attractive delta potentials supported by any closed Jordan curve in $\mathbb{R}^2$ as the coupling constant tends to infinity has been studied in \cite{ExnerYoshitomi2002} and the asymptotic expansions of the bound state energies are found for two attractive delta potentials supported by two concentric circles as their distance tends to zero as well as to  infinity in \cite{Kondej2016}.  
\end{myremark}

\subsection{Perturbative First Order Stationary Scattering Problem}
\label{First Order Stationary Scattering Problem for deformed Circle}

The function $\tilde{\Phi}$ can be analytically continued onto the complex plane using (\ref{deformedcirclePhiforboundstate}) and $\tilde{\Phi}(E_k+i0)$ can be evaluated in terms of the variable $k>0$
\begin{eqnarray} & & \hskip-1cm 
 \tilde{\Phi}(E_k+i0) = \frac{1}{\lambda}- \frac{i}{4} J_0(kR) H_{0}^{(1)}(kR) \nonumber \\ & & \hspace{3cm} + \, \frac{\epsilon}{2\pi^2}  \left(-\frac{1}{2R} + \frac{i \pi k}{2} J_0(kR) H_{1}^{(1)}(kR) \right) \left(\int_{0}^{2\pi} h(\theta) d \theta \right) + O(\epsilon^2) \;. \label{Phideformedcirclescattering} 
\end{eqnarray}
Let $\theta'$ be the angle between $\mathbf{k}'$ and $\mathbf{k}$, which is the momentum of the incoming particle chosen to be parallel to the $x$ axis for simplicity. Then, we get
\begin{eqnarray} & & \hskip-1cm 
\langle \mathbf{k}' | \tilde{\Gamma} \rangle = \left(1 + \frac{\epsilon}{2\pi R}  \int_{0}^{2\pi} h(\theta) d \theta \right) \bigg(J_0(kR) -\frac{\epsilon}{2\pi R} \int_{0}^{2\pi} e^{-ik R \cos(\theta-\theta')} h(\theta) d \theta \nonumber \\ & & \hspace{4cm}  - \, \frac{i k \epsilon}{2\pi} \int_{0}^{2\pi} e^{-i k R \cos(\theta-\theta')} \cos(\theta-\theta') h(\theta) d \theta \bigg) + O(\epsilon^2) \;. \label{k'gamma}
\end{eqnarray}
Hence, using the above results (\ref{Phideformedcirclescattering}) and (\ref{k'gamma}), and the formula for the scattering amplitude 
$\tilde{f}(\mathbf{k} \rightarrow \mathbf{k}') = \frac{1}{4} \sqrt{\frac{2}{\pi k}} \langle \mathbf{k}' | \tilde{\Gamma} \rangle (\tilde{\Phi}(E_k+i0))^{-1} \langle \tilde{\Gamma} | \mathbf{k} \rangle$ we get

\begin{mytheo} Under the small deformation of the circle described by (\ref{deformationofcurve}), the scattering amplitude up to first order in $\epsilon$ is given by  
\begin{eqnarray} & & \hskip-0.5cm
\tilde{f}(\mathbf{k} \rightarrow \mathbf{k}') = \frac{1}{4}\sqrt{\frac{2}{\pi k}}\Big( \frac{1}{\lambda}-
\frac{i}{4} J_0(k R) H_0^{(1)} (k R)\Big)^{-1} \nonumber \\ & & \times \Bigg[ J_0^2(k R) + \epsilon \Bigg( \frac{2}{R} J_0^2(k R) \langle h \rangle -J_0(k R)\int_0^{2\pi} \left[g(\theta-\theta')+g^*(\theta)\right] h(\theta) \; \frac{d\theta}{2\pi R}  \nonumber\\ 
& & \hspace{0.5cm } + \, J_0^2(k R)\Big( \frac{1}{\lambda}-
\frac{i}{4} J_0(k R) H_0^{(1)} (k R)\Big)^{-1}\Big(\frac{1}{2\pi R}-\frac{i k}{2}J_0(k R)H_1^{(1)}(k R)\Big) \langle h \rangle \Bigg)\Bigg] + O(\epsilon^2)\;,
\end{eqnarray}
where  we introduce a function
$g(\phi)=e^{-ikR\cos(\phi)}[1+ikR\cos(\phi)]$ to simplify our expressions.
\end{mytheo}

Note that we have an expansion in the form $\tilde f=\tilde f^{(0)}+\epsilon \tilde f^{(1)}$ here, that implies the total scattering cross section can be found as
\begin{equation}
    \sigma(\mathbf{k})=\int_{0}^{2\pi}  |\tilde f^{(0)}|^2 \; d\theta +\epsilon \int_{0}^{2\pi} \left[ (\tilde f^{(0)})^* \tilde f^{(1)} + \tilde f^{(0)} (\tilde f^{(1)})^* \right] \; d\theta +O(\epsilon^2).
\end{equation}

\begin{figure}
    \centering
    \includegraphics{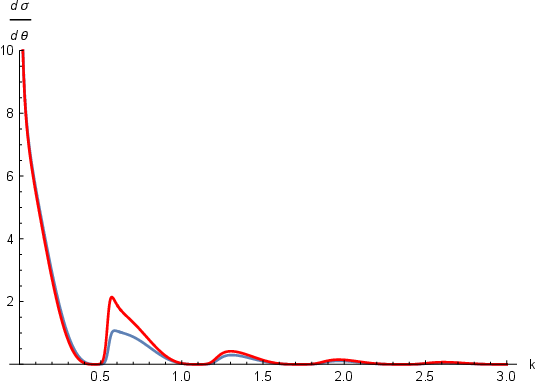}
    \caption{Differential cross sections as a function of $k$ from a circular defect and deformed circular defect (red curve) for $h(\theta)=\sin^2 \theta$, and $R=5$, $\lambda=40$, and $\epsilon=0.1$ units.}
    \label{fig:diffcrosssectiondeformedcircle}
\end{figure}

The differential cross sections as a function of $k$ for the circular defect and deformed circular defect for a particular deformation $h(\theta)= h_0 \sin^2 \theta$ is plotted in Fig. \ref{fig:diffcrosssectiondeformedcircle}, with $h_0=1$ unit.

\section{Small Deformations of a Sphere}
\label{Small Deformations of a Sphere}

We consider a particular regular surface, a sphere $S^2$ centered at the origin with radius $R$. Let $\boldsymbol{\sigma}:(0,\pi)\times (0,2\pi) \rightarrow S^2$ be a local chart, given by (\ref{localchartsphere}). Suppose that $\tilde{\Sigma}$ is the small deformation of the sphere along its normal direction, defined by
\begin{eqnarray}
\boldsymbol{\tilde{\sigma}}(\theta, \phi):= \boldsymbol{\sigma}(\theta, \phi) + \epsilon h(\theta, \phi) \mathbf{N}(\theta, \phi) \;, \label{deformedshpere}
\end{eqnarray}
where $\epsilon$ is a small deformation parameter, $\mathbf{N}$ is the normal vector field on the sphere, and $h$ is a smooth function on the sphere. If $|\epsilon|$ is sufficiently small, it is well-known that the deformed sphere $\tilde{\Sigma}$ is a regular surface \cite{bar2010elementary} and its surface area up to order $\epsilon$ is given by
\begin{eqnarray}
A(\tilde{\Sigma}) = A(\Sigma) - 2 \epsilon  \int_{0}^{2\pi} \int_{0}^{\pi}  H(\theta, \phi) h(\theta, \phi) R^2 \sin \theta d \theta d \phi +O(\epsilon^2) \;, \label{deformationofarea}
\end{eqnarray}
where $H=1/R$ is the mean curvature of the sphere. To simplify the notation, we will write $d \Omega$ instead of $\sin \theta d \theta d \phi$,  and $\Omega$ as the argument of the functions on the sphere.

\subsection{Perturbative First Order Calculation of the Bound State Energy}
\label{First Order Calculation of the Bound State Energy for Deformed Sphere}

We consider here that the interaction is formally represented by $\lambda |\tilde{\Sigma} \rangle \langle \tilde{\Sigma}|$ - rank one perturbation   supported on a deformed sphere.
The resolvent can be similarly constructed for the deformed spherical defect by following the same line of arguments discussed above. The explicit form of the resolvent operator is given by
\begin{eqnarray}
R(E)= R_0(E) + R_{0}(E) |\tilde{\Sigma} \rangle \tilde{\Phi}^{-1}(E) \langle \tilde{\Sigma}| R_0(E)\;,
\end{eqnarray}
where
\begin{eqnarray}
\tilde{\Phi}(E)= \frac{1}{\lambda} -  \langle \Tilde{\Sigma} | R_0(E) | \tilde{\Sigma} \rangle \;. \label{Phideformedsphere}
\end{eqnarray}

For this part, we  assume that the sphere problem has a bound state solution.
We will choose $E=-\nu^2$, as we will be interested in a bound state to begin with.
If we use the realization in the  Fourier domain, the resolvent kernel is given by
\begin{eqnarray}
R_0(\mathbf{r}, \mathbf{r'}|-\nu^2)= \int_{\mathbb{R}^3} \frac{e^{i \mathbf{p}\cdot (\mathbf{r}-\mathbf{r'})}}{p^2 +\nu^2} \frac{d^3 p}{(2\pi)^3} \label{resolventkernel}
\end{eqnarray}
Our aim is to calculate the function $\tilde{\Phi}(-\nu^2)$ up to order $\epsilon$. Using (\ref{deformationofarea}) and expanding the terms up to order $\epsilon$, we have
\begin{eqnarray} & & \tilde{\Phi}(-\nu^2)= \frac{1}{\lambda} -
\frac{1}{(4\pi)^2} \left(1+\frac{\epsilon}{\pi R} \int_{S^2} h(\Omega) d \Omega \right) \nonumber \\  & & \hspace{1cm} \times \int_{S^{2}\times S^2}  R_0 \left(\boldsymbol{\tilde{\sigma}}(\Omega), \boldsymbol{\tilde{\sigma}}(\Omega')|-\nu^2 \right) 
\left(1-\frac{2 \epsilon}{R} \left(h(\Omega)+h(\Omega')\right)\right) d\Omega d\Omega' + O(\epsilon^2) \;. \label{Phitildedeformedsphere1}
\end{eqnarray}
The resolvent kernel up to order $\epsilon$ can be calculated using (\ref{resolventkernel})
\begin{eqnarray} & & 
R_0 \left(\boldsymbol{\tilde{\sigma}}(\Omega), \boldsymbol{\tilde{\sigma}}(\Omega')|-\nu^2 \right) \nonumber \\ & & =  \int_{\mathbb{R}^3} e^{i \mathbf{p} \cdot \mathbf{\boldsymbol{\sigma}}(\Omega)}  e^{-i \mathbf{p} \cdot \mathbf{\boldsymbol{\sigma}}(\Omega')} \frac{(1+ \epsilon (i \mathbf{p} \cdot (h(\Omega) \mathbf{N}(\Omega)-h(\Omega') \mathbf{N}(\Omega')))}{p^2 +\nu^2} \frac{d^3 p}{(2\pi)^3} +  O(\epsilon^2) \;.
\end{eqnarray}
Substituting this into (\ref{Phitildedeformedsphere1}), and keeping the first order terms in $\epsilon$ for the surface integrals of the resolvent kernel, we obtain 
\begin{eqnarray}
& & \tilde{\Phi}(-\nu^2)= \frac{1}{\lambda} -
\frac{1}{(4\pi)^2} \left(1+\frac{\epsilon}{\pi R} \int_{S^2} h(\Omega) d \Omega \right) \Bigg[ \int_{\mathbb{R}^3} \bigg( \int_{S^2 \times S^2} e^{i \mathbf{p} \cdot (\mathbf{\boldsymbol{\sigma}}(\Omega)- \mathbf{\boldsymbol{\sigma}}(\Omega'))}  d\Omega d \Omega' \bigg)  \nonumber \\ & & \times \, \frac{1}{p^2 +\nu^2} \frac{d^3 p}{(2\pi)^3} + \epsilon \bigg( 2 \int_{\mathbb{R}^3} \bigg( \int_{S^2 \times S^2} e^{i \mathbf{p} \cdot (\mathbf{\boldsymbol{\sigma}}(\Omega)- \mathbf{\boldsymbol{\sigma}}(\Omega'))} (i \mathbf{p} \cdot \mathbf{N}(\Omega)) \, h(\Omega) \; d\Omega d \Omega' \bigg) \frac{1}{p^2 +\nu^2} \frac{d^3 p}{(2\pi)^3} \nonumber \\ & & \hspace{1cm} - \, \frac{4}{R} \int_{\mathbb{R}^3} \bigg( \int_{S^2 \times S^2} e^{i \mathbf{p} \cdot (\mathbf{\boldsymbol{\sigma}}(\Omega)- \mathbf{\boldsymbol{\sigma}}(\Omega'))} h(\Omega) \; d\Omega d \Omega' \bigg) \frac{1}{p^2 +\nu^2} \frac{d^3 p}{(2\pi)^3} \bigg) \Bigg] +O(\epsilon^2)\;.
\end{eqnarray}
We have already computed the above first integral in evaluating the second diagonal element of the matrix $\Phi$ in equation (\ref{Phispherepointnew}) and the result can be expressed as
\begin{eqnarray} \langle \Sigma | R_0(-\nu^2)|\Sigma \rangle & = & 
\frac{1}{(4\pi)^2} \int_{\mathbb{R}^3} \bigg( \int_{S^2 \times S^2} e^{i \mathbf{p} \cdot (\mathbf{\boldsymbol{\sigma}}(\Omega)- \mathbf{\boldsymbol{\sigma}}(\Omega'))}  d\Omega d \Omega' \bigg) \frac{1}{p^2 +\nu^2} \frac{d^3 p}{(2\pi)^3} \nonumber \\ & = & \frac{1}{4 \pi R} K_{1/2}(\nu R) I_{1/2}(\nu R) \;. \label{Phideformedsphere1stintegral}
\end{eqnarray}
For the second integral, we will use the identity $\left(i\mathbf{p}\cdot \mathbf{N}(\Omega) \right) e^{i\mathbf{p}\cdot \boldsymbol{\sigma}(\Omega)} = \frac{\partial}{\partial R} e^{i\mathbf{p}\cdot \boldsymbol{\sigma}(\Omega)}$.  The exponential factors can be expressed in terms of the spherical Bessel functions of first kind and spherical harmonics using  the well-known expansion of the plane waves into spherical harmonics \cite{stone2009mathematics}:
\begin{eqnarray}
e^{i\mathbf{p}\cdot \boldsymbol{\sigma}(\Omega)}=4\pi \sum_{l=0}^{\infty} \sum_{m=-l}^{l} i^l j_l(p R) Y_{lm}^* (\Omega_p)Y_{lm} (\Omega) \;. \label{planewaveexpansion}
\end{eqnarray}
Here $\Omega_p$ and $\Omega$ are the polar angles of the vector $\mathbf{p}$ and $\boldsymbol{\sigma}$, respectively. Hence, we obtain
\begin{eqnarray} & & 
\int_{\mathbb{R}^3} \bigg( \int_{S^2 \times S^2} e^{i \mathbf{p} \cdot (\mathbf{\boldsymbol{\sigma}}(\Omega)- \mathbf{\boldsymbol{\sigma}}(\Omega'))} (i \mathbf{p} \cdot \mathbf{N}(\Omega)) \, h(\Omega) \; d\Omega d \Omega' \bigg) \frac{1}{p^2 +\nu^2} \frac{d^3 p}{(2\pi)^3} \nonumber \\ 
& & = (4\pi)^2 \int_{0}^{\infty} \int_{S^2} \bigg( \int_{S^2 \times S^2} \sum_{l=0}^{\infty} \sum_{m=-l}^{l} i^l \frac{\partial j_l(p R)}{\partial R} Y_{lm}^* (\Omega_p) Y_{lm} (\Omega) h(\Omega)  \nonumber \\ & & \hspace{2cm} \times  \sum_{l'=0}^{\infty} \sum_{m'=-l}^{l} (-i)^{l'} \frac{\partial j_{l'}(p R)}{\partial R} Y_{l'm'} (\Omega_p)Y_{l'm'}^{*}(\Omega') d \Omega d \Omega' \bigg)  \frac{d \Omega_p p^2 d p }{(2\pi)^3} \;.
\end{eqnarray}
By the orthonormality of the spherical harmonics $\int_{S^2} Y_{lm}(\Omega) Y_{l'm'}(\Omega) d\Omega=\delta_{ll'} \delta_{mm'}$,
integrations over $\Omega_p$ and $\Omega'$ lead to
\begin{eqnarray} & & 
\int_{\mathbb{R}^3} \bigg( \int_{S^2 \times S^2} e^{i \mathbf{p} \cdot (\mathbf{\boldsymbol{\sigma}}(\Omega)- \mathbf{\boldsymbol{\sigma}}(\Omega'))} (i \mathbf{p} \cdot \mathbf{N}(\Omega)) \, h(\Omega) \; d\Omega d \Omega' \bigg) \frac{1}{p^2 +\nu^2} \frac{d^3 p}{(2\pi)^3} \nonumber \\ 
& & =\frac{(4\pi)^2}{(2\pi)^3} \int_{0}^{\infty}  j_0(p R) (-j_1(p R)) \frac{p^3}{p^2+\nu^2} d p \left( \int_{S^2} h(\Omega) d \Omega \right) \;,
\end{eqnarray}
where we have used $Y_{00}(\Omega)=1/\sqrt{4\pi}$ and the relation $\frac{d j_0(x)}{d x}=-j_1(x)$. We now use $j_l(x)=\sqrt{\frac{\pi}{2x}} J_{l+1/2} (x)$
and decompose $\frac{p^2}{p^2+\nu^2}$ as $1-\frac{\nu^2}{p^2+\nu^2}$ together with the formulas (6.512) and (6.577) in \cite{gradshteyn2014table} for the integrals of the Bessel functions 
\begin{eqnarray}
\int_{0}^{\infty} J_{1/2}(p R) J_{3/2}(p R) dp & = & \frac{1}{2R} \;, \\ \int_0^\infty  J_{3/2} (p R) J_{1/2}(p R) \frac{d p}{p^2+\nu^2} & = & \frac{1}{\nu} I_{3/2}(\nu R) K_{1/2}(\nu R) \;,
\end{eqnarray}
to get
\begin{eqnarray} & & \hskip-3cm
\int_{\mathbb{R}^3} \bigg( \int_{S^2 \times S^2} e^{i \mathbf{p} \cdot (\mathbf{\boldsymbol{\sigma}}(\Omega)- \mathbf{\boldsymbol{\sigma}}(\Omega'))} (i \mathbf{p} \cdot \mathbf{N}(\Omega)) \, h(\Omega) \; d\Omega d \Omega' \bigg) \frac{1}{p^2 +\nu^2} \frac{d^3 p}{(2\pi)^3} \nonumber \\ 
& & =- \frac{1}{R}  \left( \int_{S^2} h(\Omega) d \Omega \right) \left( \frac{1}{2R}- \nu K_{1/2}(\nu R) I_{3/2}(\nu R) \right) \;.
\label{Phideformedsphere2ndintegral}
\end{eqnarray}
By applying similar arguments, we can find easily the last integral
\begin{eqnarray} & & \hskip-3cm
\int_{\mathbb{R}^3} \bigg( \int_{S^2 \times S^2} e^{i \mathbf{k} \cdot (\mathbf{\boldsymbol{\sigma}}(\Omega)- \mathbf{\boldsymbol{\sigma}}(\Omega'))} h(\Omega) \; d\Omega d \Omega' \bigg) \frac{1}{k^2 +\nu^2} \frac{d^3 k}{(2\pi)^3} \nonumber \\ & & \hspace{3cm} = \frac{2}{R} K_{1/2}(\nu R) I_{1/2}(\nu R) \left( \int_{S^2} h(\Omega) d\Omega \right) \;.  \label{Phideformedsphere3thintegral}
\end{eqnarray}
Combining all these results (\ref{Phideformedsphere1stintegral}), (\ref{Phideformedsphere2ndintegral}) and (\ref{Phideformedsphere3thintegral}), we obtain
\begin{eqnarray} & & \hskip-3cm
\tilde{\Phi}(-\nu^2) = \frac{1}{\lambda} - \frac{1}{4\pi R} I_{1/2}(\nu R) K_{1/2}(\nu R) \nonumber \\ \hspace{3cm} & & + \, \frac{\epsilon}{8 \pi^2 R} \left(- \frac{1}{2R} + \nu I_{1/2}(\nu R) K_{3/2}(\nu R) \right) \left(\int_{S^2} h(\Omega) d \Omega \right) \;, \label{deformedspherephi}
\end{eqnarray}
where we have used $I_{1/2}(x)K_{3/2}(x)+I_{3/2}(x)K_{1/2}(x)=1/x$.

It is important to notice that the formula for the function $\tilde{\Phi}$ is very similar to the one obtained for the deformed circular defect case, however there is a difference. The eigenvalue flow can be obtained again by writing  $I_{1/2}(\nu R) K_{1/2}(\nu R)$ as (from the formula (6.577) in \cite{gradshteyn2014table}):
\begin{eqnarray}
\frac{1}{\lambda} = \frac{1}{4\pi R}  I_{1/2}(\nu R) K_{1/2}(\nu R) = \frac{1}{4\pi R} \int_{0}^{\infty} \frac{x}{x^2+\nu^2 R^2} J_{1/2}^{2}(x) d x \;.
\end{eqnarray}
As one can see, the right hand side of the above equation is a decreasing function of $\nu$ for given parameters $\lambda$ and $R$. Yet the product $I_{1/2}(\nu R) K_{1/2}(\nu R)$ is finite as $\nu\to 0^+$, so  there may not always be a solution if $\lambda$ is small enough. If there is a solution then it is unique,  say  $\nu_0$, to the above equation. {\it We assume that this is the case}. 

Let $\nu= \nu_0 + \epsilon \nu_1 + O(\epsilon^2)$, then the bound state energy up to order $\epsilon$ can  be found by solving the zeroes of  $\tilde{\Phi}$ by expanding terms around $\nu=\nu_0$. Hence, we find   

\begin{mytheo} Under the small deformation of the sphere described by (\ref{deformedshpere}), the bound state energy of the system up to the first order in $\epsilon$ is given by  
\begin{eqnarray} & & \hskip-1cm
E_B \nonumber  = -\nu_{0}^{2}  -  \epsilon \, \nu_{0}^{2}\; \left( \frac{\frac{1}{2 \nu_0 R} -  I_{1/2}(\nu_0 R) K_{3/2}(\nu_0 R)}{I_{3/2}(\nu_0 R) K_{1/2}(\nu_0 R) - I_{1/2}(\nu_0 R) K_{3/2}(\nu_0 R) + \frac{1}{\nu_0 R} I_{1/2}(\nu_0 R) K_{1/2}(\nu_0 R) }\right)\nonumber \\ & & \hspace{5cm} \times \, \left( \frac{1}{\pi R} \int_{S^2} h(\Omega) d \Omega \right)   + O(\epsilon^2) \;.  
\end{eqnarray}
\end{mytheo}

Not surprisingly, {\it this  result has the same geometric interpretation as in the case of circle}, we replace the original sphere with another sphere 
of slightly different radius $R-\epsilon \langle h \rangle$, with $\langle h \rangle=\frac{1}{4\pi R^2}\int_{S^2} h(\Omega) R^2 d\Omega $  and then look for the small change in the energy because  of this alteration, as a result of this computation, we recover the above expression. Hence,
\begin{mycorollary}
A small deformation in the normal direction  of a given sphere, which supports an attractive delta function, leads to a perturbation of the original  bound state energy,  to   first order the resulting change can be obtained as follows:  increase the initial radius  by  an amount equal to the average of the deformation over the given sphere,  then compute  the first order perturbation of the  bound state  energy corresponding to  this new sphere with the same coupling constant. 
\end{mycorollary}

\begin{figure}[h!]
    \centering
    \includegraphics{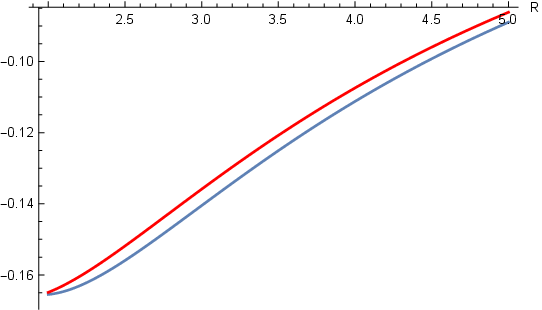}
    \caption{Bound state energy for the spherical defect and for the deformed spherical defect (red curve) versus $R$ with $\epsilon=0.1$, $\lambda=10$ units.}
    \label{fig:boundstatevsRdeformedsphere}
\end{figure}

For a particular deformation $h(\theta)=h_0 \sin \theta$ with $h_0=1$ unit, one can numerically plot how the bound state energies change with respect to $R$ for a given $\lambda$, as shown in Fig. \ref{fig:boundstatevsRdeformedsphere}.

\begin{myremark} Other types of  deformations of the support of delta potentials are introduced and  analysed in \cite{ExnerYoshitomi2003}.
In that work, the asymptotic expansions  of the eigenvalues for  the deformation of a surface in $n$ dimensions for a  Hamiltonian  given by $-\Delta- \beta \delta(\cdot-\Sigma \setminus S_{\epsilon})$, where $S_\epsilon$ is a family of measurable subsets of a chosen  surface $\Sigma$ each one of which has  size vanishing by  $O(\epsilon)$ as $\epsilon \to 0$. However, such deformations are different from the ones that we consider in the present  work. We assume small deformations of  a sphere, in the normal direction,  and study how  the bound state energies change.  
\end{myremark}

\begin{myremark}
A particular class of small and normal deformations of a sphere is the area-preserving ones. In \cite{ExnerFraas09}, if Hamiltonian for delta potential supported by a sphere satisfies a critical property, namely if it has an empty discrete spectrum and a threshold resonance,  it is then  shown that any sufficiently small smooth area preserving radial deformation leads to isolated eigenvalues. Here our aim is to look at sufficiently small smooth  radial deformations of a sphere supporting a   rank one perturbation. 
\end{myremark}

\subsection{Perturbative First Order Stationary Scattering Problem}
\label{First Order Stationary Scattering States for Deformed Sphere}

For the deformed spherical defect, the function $\tilde{\Phi}$ can be analytically continued onto the complex plane using (\ref{deformedspherephi}) and $\tilde{\Phi}(E_k+i0)$ can then be evaluated in terms of the variable $k>0$
\begin{eqnarray} & & \hskip-1cm
 \tilde{\Phi}(E_k+i0) = \frac{1}{\lambda}- \frac{i}{8 R} J_{1/2}(kR) H_{1/2}^{(1)}(kR) \nonumber \\ & & \hspace{2cm} + \,\frac{\epsilon}{8 \pi^2 R}  \left(-\frac{1}{2R} + \frac{i \pi k}{2} J_{1/2}(kR) H_{3/2}^{(1)}(kR) \right) \left(\int_{S^2} h(\Omega) d \Omega \right) +  O(\epsilon^2) \;. \label{Phideformedspherescattering} 
\end{eqnarray}
For the scattering amplitude, we need to find the expression $ \langle \tilde{\Sigma}|\mathbf{k} \rangle$ in terms of the deformation function $h(\Omega)$: 
\begin{eqnarray}
 \langle \tilde{\Sigma}|\mathbf{k} \rangle = \frac{1}{A(\tilde{\Sigma})} \int_{S^2} e^{i \mathbf{k}\cdot \boldsymbol{\tilde{\sigma}}(\Omega)}  R^2 \left(1-\frac{2 \epsilon}{R} h(\Omega)\right) \;  d \Omega \;.
\end{eqnarray}
By expanding the exponential $e^{i \epsilon h(\Omega) \mathbf{k} \cdot \mathbf{N}(\Omega)}$  in $\epsilon$ and expanding $A(\tilde{\Sigma})$ in $\epsilon$ from the formula (\ref{deformationofarea}), it is easy to show that
\begin{eqnarray} & & \hskip-2cm 
\langle \tilde{\Sigma}|\mathbf{k} \rangle = \left(1 + \frac{\epsilon}{2\pi R}  \int_{S^2} h(\Omega) d \Omega \right) \bigg(\frac{\sin (k R)}{k R} -\frac{\epsilon}{2\pi R} \int_{S^2} e^{i \mathbf{k} \cdot \boldsymbol{\sigma}(\Omega)} h(\Omega) d \Omega \nonumber \\ & & \hspace{4cm}  + \, \frac{i \epsilon}{4\pi} \int_{S^2} e^{i \mathbf{k} \cdot \boldsymbol{\sigma}(\Omega)} (\mathbf{k} \cdot \mathbf{N}(\Omega)) h(\Omega) d \Omega \bigg) + O(\epsilon^2) \;. \label{gammak}
\end{eqnarray}
For simplicity, we consider a particular class of deformations, where $h(\Omega)=h(\theta)$. In this case, let $\theta'$ be the angle between $\mathbf{k}'$ and $\mathbf{k}$, which is the momentum of the incoming particle chosen to be parallel to the $z$ axis. Then, using $\tilde{f}(\mathbf{k} \rightarrow \mathbf{k}') = \frac{1}{4\pi} \langle \mathbf{k}' | \tilde{\Sigma} \rangle (\tilde{\Phi}(E_k+i0))^{-1} \langle \tilde{\Sigma} | \mathbf{k} \rangle$ we get the explicit expression for the scattering amplitude for a given deformation $h$:

\begin{mytheo}
 Under the small deformation of the sphere described by (\ref{deformedshpere}), the scattering amplitude up to the first order in $\epsilon$ is given by  
\begin{eqnarray} & & \hskip-0.7cm \tilde{f}(\mathbf{k} \rightarrow \mathbf{k}') = \frac{1}{4 \pi} \left(\frac{1}{\lambda}- \frac{i}{8 R} J_{1/2}(kR) H_{1/2}^{(1)}(kR)\right)^{-1} \\ & & \hspace{0.5cm} \times \Bigg\{ \frac{\sin^2 k R}{k^2 R^2} + \epsilon \bigg[ \frac{4 \langle h \rangle \sin^2 k R}{k^2 R^3}- \frac{\sin k R}{k R^2} \int_{0}^{\pi} \left(g(\theta-\theta')+g^*(\theta) \right) \sin \theta \; h (\theta) \; d \theta \\ & & \hskip-0.3cm + \, \frac{\sin^2 k R}{k^2 R^2} \left( \frac{1}{4 \pi R^2}- \frac{i k}{4 R} J_{1/2}(k R) H^{(1)}_{3/2}(k R) \right) \left(\frac{1}{\lambda}- \frac{i}{8 R} J_{1/2}(kR) H_{1/2}^{(1)}(kR)\right)^{-1} \langle h \rangle \bigg] \Bigg\} + O(\epsilon^2) \;,
\end{eqnarray}
where $g(\phi):= e^{-i k R \cos \phi} \left(1+\frac{i k R}{2} \cos \phi \right)$.
\end{mytheo}

\begin{figure}
    \centering
   \includegraphics{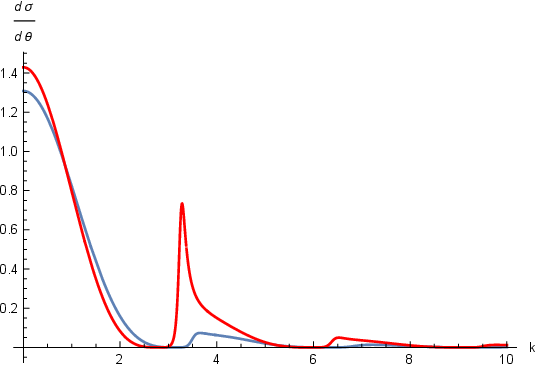}
    \caption{Differential cross sections as a function of $k$ from a spherical defect and deformed spherical defect (red curve) for $h(\theta)=\sin \theta$, and $R=1$, $\lambda=100$, and $\epsilon=0.1$ units.}
    \label{fig:diffcrosssectiondeformedsphere}
\end{figure}

The differential cross sections as a function of $k$ for the spherical defect and deformed spherical defect for a particular deformation $h(\theta)=\sin \theta$ is plotted in Fig. \ref{fig:diffcrosssectiondeformedsphere}.

\section*{Appendix: Trotter-Kato Theorem} \label{TrotterKatotheorem}

This is a slightly different version of Trotter-Kato theorem than typically found in the literature\cite{Reed1972methods}, as stated also in \cite{rajeevdimock}:  
\begin{theorem*}
Suppose that $H_n$ be a sequence of self-adjoint operators with resolvents $R_n(E)=(H_n-E)^{-1}$ defined for all complex numbers $E$ except a closed proper subset $U$ of $\mathbb{R}$. Furthermore, assume that $R_n(E)$ converges strongly for some $E \notin U$ and this limit is invertible. Then, there exists a self-adjoint operator $H$ with resolvents $R(E)=(H-E)^{-1}$ such that $R_n(E)$ converges strongly to $R(E)$ for all complex numbers $E \notin U$.   
\end{theorem*}
The idea of the proof is essentially the same as the original Trotter-Kato theorem. In our problem, we choose a sequence $\epsilon_n=1/n$. If $n$ is sufficiently large $\det \Phi(\epsilon, E)\neq 0$ if $E$ satisfies $\det \Phi(E)\neq 0$. Then, $R(\epsilon_n,E)$ is defined for all complex $E$ except a closed proper subset $U$ of $\mathbb{R}$, namely $U= \{E \in [0,\infty): \det \Phi(E)=0 \}$. Since we have shown that $R(\epsilon_n, E)$ converges strongly for some $E \notin U$ (e.g., choose $E$ to be a sufficiently large negative real number) and the limit is invertible, we conclude that there exists a self-adjoint operator $H$ with resolvents $R(E)=(H-E)^{-1}$ such that $R(\epsilon_n,E)$ converges strongly to $R(E)$ for all complex numbers $E \notin U$ thanks to the above theorem.

\section*{Acknowledgments}
O. T. Turgut would like to thank A. Michelangeli for many informative discussions on mathematical aspects of singular interactions in general as well as A. Mostafazadeh for his interest in these problems. The authors are also grateful to the anonymous referee for her/his useful remarks and suggestions.

\end{document}